\documentclass[structabstract]{aa}
\usepackage{natbib,twoopt}
\usepackage{graphicx}
\usepackage{txfonts}
\usepackage{color}
\bibpunct{(}{)}{;}{a}{}{,}             
\makeatletter
  \newcommandtwoopt{\citeads}[3][][]{\href{http://adsabs.harvard.edu/abs/#3}%
    {\def\hyper@linkstart##1##2{}%
     \let\hyper@linkend\@empty\citealp[#1][#2]{#3}}}
  \newcommandtwoopt{\citepads}[3][][]{\href{http://adsabs.harvard.edu/abs/#3}%
    {\def\hyper@linkstart##1##2{}%
     \let\hyper@linkend\@empty\citep[#1][#2]{#3}}}
  \newcommandtwoopt{\citetads}[3][][]{\href{http://adsabs.harvard.edu/abs/#3}%
    {\def\hyper@linkstart##1##2{}%
     \let\hyper@linkend\@empty\citet[#1][#2]{#3}}}
  \newcommandtwoopt{\citeyearads}[3][][]%
    {\href{http://adsabs.harvard.edu/abs/#3}
    {\def\hyper@linkstart##1##2{}%
     \let\hyper@linkend\@empty\citeyear[#1][#2]{#3}}}
\makeatother

\begin{document}

\title{Abundances of carbon-enhanced metal-poor stars as constraints on their formation\thanks{Based on observations obtained at ESO Paranal Observatory, programmes 084.D-0117(A) and 085.D-0041(A)}
}

\titlerunning{Abundances in normal and C-enhanced metal-poor stars}

\author{C.J. Hansen \inst{1,2},
B. Nordstr\"om \inst{1,3}
T.T. Hansen\inst{2},
C.R. Kennedy \inst{4,6},
V.M. Placco \inst{5},
T.C. Beers \inst{5},
J. Andersen \inst{1,3,7},
G. Cescutti \inst{8,9}
\and
C. Chiappini \inst{8}
}
\authorrunning{C.~J. Hansen et al.}
\offprints{C.~J. Hansen, \email{cjhansen@dark-cosmology.dk}}

\institute{Dark Cosmology Centre, The Niels Bohr Institute, 
Juliane Maries Vej 30, DK-2100 Copenhagen, Denmark
\email{cjhansen@dark-cosmology.dk, birgitta@nbi.ku.dk, ja@nbi.ku.dk}
\and
Zentrum f\"ur Astronomie der Universit\"at Heidelberg, Landessternwarte, K\"onigstuhl 12, D-69117 Heidelberg, Germany 
\email{thansen@lsw.uni-heidelberg.de} 
\and
Stellar Astrophysics Centre, Department of Physics and Astronomy,
Aarhus University, Ny Munkegade 120, DK-8000 Aarhus C, Denmark
\and
Research School of Astronomy and Astrophysics, Australian National University, Cotter Road, Weston, ACT 2611, Australia
\and
Department of Physics and JINA Center for the Evolution of the Elements, 
University of Notre Dame, Notre Dame, IN 46556, USA \email{tbeers@nd.edu, vplacco@nd.edu}
\and 
Department of Chemistry, Biochemistry, and Physics, University of Tampa, 401 W. Kennedy Blvd., Tampa, FL 33606, USA \email{ckennedy@ut.edu}
\and
Nordic Optical Telescope Scientific Association, Apartado 474, ES-38700 Santa Cruz de La Palma, Canarias, Spain 
\and
Leibniz-Institut f\"ur Astrophysik Potsdam (AIP), D-14482 Potsdam, Germany
\email{cristina.chiappini@aip.de}
\and
Centre for Astrophysics Research, Univeristy of Hertfordshire, Hatfield, Herts,
AL10 9AB, UK \email{g.cescutti@herts.ac.uk}
}

\date{Received August, 2015; accepted November}

\abstract{An increasing fraction of carbon-enhanced metal-poor
(CEMP) stars is found as their iron abundance, [Fe/H], decreases below
[Fe/H] $ = -2.0$. The CEMP-$s$ stars have the highest absolute carbon 
abundances, [C/H], and are thought to owe their enrichment in carbon and the slow 
neutron-capture ($s$-process) elements to mass transfer from a former 
asymptotic giant-branch (AGB) binary companion. The most Fe-poor CEMP stars 
are normally single, exhibit somewhat lower [C/H] than CEMP-$s$ stars, 
but show no $s$-process element enhancement (CEMP-no stars). CNO abundance 
determinations offer clues to their formation sites. }
{Our aim is to use the
medium-resolution spectrograph X-shooter/VLT to determine stellar
parameters and abundances for C, N, Sr, and Ba in several classes of
CEMP stars, in order to further classify and constrain the astrophysical
formation sites of these stars.} 
{Atmospheric parameters for our programme
stars were estimated from a combination of $V-K$ photometry, model
isochrone fits, and estimates from a modified version of the SDSS/SEGUE
spectroscopic pipeline. We then used X-shooter spectra in conjunction
with the 1D LTE spectrum synthesis code MOOG, and 1D ATLAS9 atmosphere 
models to derive stellar abundances, and, where possible, isotopic
$^{12}$C/$^{13}$C ratios.}
{C, N, Sr, and Ba abundances (or limits) are derived for a sample 
of 27 faint metal-poor stars for which the X-shooter spectra have sufficient 
S/N ratios. These moderate resolution, low S/N ($\sim
10-40$) spectra prove sufficient to perform limited chemical tagging 
and enable assignment of these stars into the CEMP sub-classes
(CEMP-$s$ and CEMP-no). According to the derived abundances, 17 of our sample stars are CEMP-$s$ and three are CEMP-no, while the remaining seven are carbon-normal. For four CEMP stars, the sub-classification remains uncertain, and two of them may be pulsating AGB stars.}
{The derived stellar abundances trace the formation
processes and sites of our sample stars. The [C/N] abundance ratio is useful to identify stars with chemical 
compositions unaffected by internal mixing, and the [Sr/Ba] abundance ratio 
allows us to distinguish between CEMP-$s$ stars with AGB progenitors and 
the CEMP-no stars. Suggested formation sites for the latter include 
faint supernovae with mixing and fallback and/or primordial, rapidly-rotating, 
massive stars (spinstars). X-shooter spectra have thus proved to be valuable 
tools in the continued search for their origin.}
\maketitle

\section{Introduction}

The class of very metal-poor (VMP; [Fe/H] $< -2.0$) stars with strong
molecular absorption features of carbon, in particular the CH $G$-band,
are collectively referred to as carbon-enhanced metal-poor (CEMP) stars.
They are defined as stars with [Fe/H] $\leq -$2 and [C/Fe] $> 0.7$,
following \citet{Beers2005} and \citet{Aoki2007}, respectively. For convenience, we employ
the term ``carbonicity'' for the carbon-to-iron ratio, [C/Fe],  to
distinguish it from the absolute carbon abundance, [C/H]
\citep{Placco2011}.

In the last two decades it has been recognized that roughly 20\% of VMP
halo stars exhibit carbonicities up to several orders of magnitude
larger than Solar \citep{Marsteller2005,Rossi2005,Lucatello2006},
rising to 30\% for stars with [Fe/H] $< -3.0$, 40\% for [Fe/H] $< -3.5$,
and 75\% for [Fe/H] $< -4.0$. The handful of stars known with [Fe/H] $<
-5.0$ all exhibit large carbonicities \citep{Lee2013,Placco2014b,
Frebel2015}, including the most iron-poor star presently known,
\object{SMSS~J0313-6708}, with [Fe/H] $< -7.8$ \citep{Keller2014,
Bessel2015}. A definitive interpretation of this increasing frequency
has not yet been found, but it has been argued \citep[e.g.,
][]{Carollo2012} that CEMP stars are more frequently associated with the
outer-halo population of the Galaxy than with the inner-halo population,
suggesting differences in the nucleosynthetic production sites of carbon
in these components. \citet{Carollo2014} presented indications of the
CEMP-$s$ stars being preferentially associated with the inner-halo
population, while the CEMP-no stars are associated with the outer-halo
population. Based on these, and other recent results \citep[see, e.g.,
][]{Ito2013,Placco2014a,THansen2015}, it is becoming
increasingly clear that the CEMP-no stars may well be bona fide
second-generation stars, born from an ISM polluted with the
nucleosynthetic products of the very first stars.   

A few exceptions to this empirical rule have been presented in
\citet{Bonifacio2015} and \citet{Caffau2011}. In the former study, a star
with an upper limit of [Fe/H] $< -5.0$ was found to have a low absolute C
abundance, while the latter study presented a truly metal-poor star (SDSS
J102918+172927), with metallicity slightly above [Fe/H] = $-5.0$ but
without large carbonicity ([C/Fe] $\leq 0.9$).

In any event, most results indicate that significant amounts of carbon
were produced already at the very earliest stages of the chemical
evolution of the Milky Way and the Universe itself. The Galactic
chemical evolution (GCE) models by \citet{Cescutti2013} can account for
a large amount of the star-to-star scatter in [Sr,Ba/Fe] found at the
lowest [Fe/H], and can provide good explanations for the chemistry of
most CEMP-no stars, based on a spinstar production scenario
\citep[see][]{Cescutti2010}. However, these models alone cannot
simultaneously explain the large excess of C, N, and $s$-process
elements found in the CEMP-$s$ stars. 

High-resolution spectroscopic studies have shown that $\sim$80\% of the
known CEMP stars are CEMP-$s$ stars \citep[][based on their sample of 26 stars]{Aoki2007}. The favoured
mechanism to account for these stars is local transfer of carbon-rich
material from the envelope of an asymptotic giant-branch (AGB) star to
the surface of a surviving binary companion \citep[e.g.,][]{Herwig2005,
Bisterzo2012}. Correspondingly, detailed models for the chemical
composition of a number of CEMP-$s$ stars have been developed
\citep[e.g.,][]{Placco2013,Abate2015,Placco2015}, but no definitive proof yet
exists that they are {\em all} members of binary systems (although
a great majority are clearly binaries, see, e.g.,
\citealt{Lucatello2005, Starkenburg2014}). If a given
CEMP-$s$ star is not a member of a binary
system, eliminating the possibility of mass transfer from an evolved
companion, the excess C must have been produced by a distant external
source and implanted in the natal cloud of the star observed today. 

CEMP-no stars are known to be particularly prevalent among the
lowest-metallicity stars \citep{Aoki2007}. Indeed, of the nine stars
known with [Fe/H] $\leq -4.5$, seven are CEMP-no stars
\citep{Placco2014b,Bonifacio2015,Frebel2015}, suggesting that C was produced and
enriched already in the very first stellar generations. Their low binary 
frequency (\citet{Starkenburg2014}; Hansen et al. 2015b, subm.), $\sim 17 \pm 5$\%, consistent
with that of halo stars with normal carbon content, and their lack of
$s$-process-element abundance signatures makes a local AGB binary
companion an unlikely source of their C excess, so another production site
must be found.

One possible progenitor of the CEMP-no stars are massive, rapidly
rotating, mega metal-poor ([Fe/H] $< -6.0$) stars \citep{Meynet2006,
Hirschi2007,Frischknecht2012,Maeder2015}, the so-called ``spinstars''. Another
suggested production site for the material incorporated into CEMP-no
stars are the so-called ``faint supernovae'' associated with the first
generations of stars, which experience extensive mixing and fallback
during their explosions \citep[e.g.,][]{Umeda2003, Tominaga2007, Nomoto2013,
Tominaga2014}.   It remains possible that both of these suggested
progenitors may have played a role.

It is noteworthy that extremely metal-poor ([Fe/H] $\sim -3.0$) damped
Lyman-$\alpha$ (DLA) systems at high redshift ($z = 2-3$) with enhanced
[C/Fe] (and other light elements) were recently reported by
\citet{Cooke2011, Cooke2012}, indicating the existence of a C-enhanced
ISM at very early times. \citet{Kobayashi2011} associate C and other
elemental abundance signatures of CEMP-no stars with production by faint
supernovae, and \citet{Matsuoka2011} have also argued for a strong
carbon production in the early Universe, based on their analysis of the
very distant radio galaxy \object{TN~0924-2201} ($z = 5.19$)
\footnote{Carbon ([CII] is also measured at even higher redshifts (z$\sim7.1$) 
in a quasar (\object{J112001.48+064124.3}), and is found to be lower than
in other quasars at z$\sim6$ \citep{Venemans2012}. The carbon-to-far
infrared flux ratio indicate the presence of a significant amount of
cold gas and dust in the early Universe.}.

Progress on these issues requires as complete an inventory of the most
important elemental- and isotopic-abundance ratios for CEMP stars as
possible, in particular C, N, Fe, Sr, Ba, and $^{12}$C/$^{13}$C. The
highly efficient X-shooter instrument on the ESO VLT, covering the full
wavelength range from 300 nm to 2.3 $\mu$m range in a single exposure,
allows us to measure the NH band, the CH band, Sr, and Ba in the near-UV
to the optical and near-IR. 

This paper reports our abundance analysis results for a sample of
27 MP or VMP stars, 20 of which are CEMP stars, and is outlined as follows.
Section 2 discusses the target selection for our study. Details of our
X-shooter observations are given in Sect. 3, while Sect. 4 describes our
derivation of elemental abundances for the programme stars. Section 5
presents our results, while Sect. 6 offers a discussion of how our key
results advance our ultimate goal of understanding the origin of CEMP
stars. Finally, our conclusions are presented in Sect. 7.

\section{Target selection}

Our targets were selected from the ``Catalogue of carbon stars
found in the Hamburg-ESO survey'' \citep{Christlieb2001} and ``Bright
Metal-poor Stars from the Hamburg-ESO Survey'' \citep{Frebel2006}. For
this study we selected the brightest VMP stars, all of which are
subgiants or giants, along with a number of additional chemically
``normal'' metal-poor stars. Initial estimates of atmospheric parameters
for most of our candidates were available from application of a modified
version of the SEGUE Stellar Parameter Pipeline (n-SSPP), described in
more detail below. Based on these, we attempted to select objects in a
temperature range suitable for estimation of molecular abundances such
as those from molecular C and N bands. The X-shooter spectrograph
remains efficient even at the NH band at 336 nm, so targets with an
effective temperature down to $\sim$4000 K could be observed to a
S/N ratio sufficient for abundance analysis in faint stars with
reasonable integration times. 

The list of our programme stars is given in Table\ref{table:obs},
together with the integration times used and the S/N ratios obtained.
The $B$ and $V$ magnitudes listed are from \citet{Beers2007} unless
otherwise stated, while $K$ (= $K_s$) and $J$ magnitudes are from the
2MASS catalogue \citep{Cutri2003}. After our observations were already
completed, the star \object{HE~0430$-$1609} was found to be a single-lined
spectroscopic binary with a period of the order of 3 years
(Hansen et al. 2015c, subm.); this should not affect our results, as the low S/N we
obtained for this star already yields larger uncertainties.

Our aim was to obtain spectra with sufficient S/N ratio around the NH
band at 336nm, guaranteeing a higher S/N in the redder parts of the
wavelength range of X-shooter. However, measuring an actual S/N ratio
from the spectra of cool stars in the near-UV is difficult, since the region
is extremely crowded with molecular lines, and a clean continuum often
cannot be defined. We have therefore measured the S/N in a region near
400\,nm, which still contains a substantial number of (mostly)
molecular lines (see Table\ref{table:obs}). This S/N can be taken as
representative of the spectrum quality near the Sr line that we
employed.


\begin{table*}
\centering
\begin{minipage}{150mm}
\caption{Coordinates, photometry, integration times, and resulting S/N ratios
for our programme stars}
\label{table:obs} 
\begin{tabular}{lcrllrrrccrrrrl}
\hline

         Name            & RA        &   Dec   &  $B$ &  $V$   & $J$ &  $K$   & Int. Time [s]&  S/N $@ 4000$\AA \\
\hline    
  
          \object{HE~0058$-$3449} &    01:01:21.6 &     -34:33:11  &   13.91  &   13.20         & 11.78  &   11.33  & 380 & 13   \\
	    \object{HE~0206$-$1916} &    02:09:19.6  &    -19:01:56  &   15.12  &   14.00         & 12.24  &   11.66  & 500 & 7   \\
          \object{HE~0241$-$3512} &    02:43 26.7  &    -35:00:12 &    15.24  &   13.91         & 12.04  &   11.40  & 500 &  8  \\
          \object{HE~0400$-$2030} &    04:02:14.8  &    -20:21:53 &    14.92  &   14.03         & 12.49  &   11.98  & 400 & 13 \\
          \object{HE~0408$-$1733} &    04:11:06.0  &    -17:25:40  &   13.51  &   12.34         & 10.37  &    9.67  & 130 & 6  \\
          \object{HE~0414$-$0343} &    04:17:16.4  &    -03:36:31 &    12.25  &   11.05         & 9.19   &    8.51  & 30  & 8  \\
          \object{HE~0430$-$1609}$^{SB}$ &  04:32:50.6  &    -16:03:38 & 14.41$^a$ &   13.21$^a$     & 11.39  &   10.74  & 230 & 6  \\
          \object{HE~0430$-$4901} &    04:31:31.1  &    -48:54:42 &    15.16  &   14.57         & 13.31  &   12.81  & 250 & 36   \\
          \object{HE~0440$-$3426} &    04:42:08.1 &     -34:21:13 &    12.59  &   11.42         &  9.65  &    8.97  & 60  & 10 \\
          \object{HE~0448$-$4806} &    04:49:33.0  &    -48:01:08 &    13.39  &   12.78         & 11.59  &   11.21  & 400 & 17 \\
          \object{HE~0516$-$2515} &    05:18:09.4 &     -25:12:25 & 15.44$^a$ &   13.66$^a$     & 11.25  &   10.35  & 600 & 10  \\
          \object{HE~1238$-$0836} &    12:41:02.4  &    -08:53:06 & 13.56$^A$ &   11.79$^A$     &  9.19  &    8.15  & 400 & 9  \\
          \object{HE~1315$-$2035} &    13:17:57.4  &    -20:50:53 &    16.81  &   15.43         & 13.68  &   13.05  & 1100 & 11\\
          \object{HE~1418$+$0150} &    14:21:01.2 &      01:37:18 & 14.05$^a$ &   12.33$^a$     &  9.99  &    9.13  & 600 & 9\\
          \object{HE~1430$-$0919} &    14:33:12.9  &    -09:32:53  &   15.68  &   14.31         & 12.48  &   11.86  & 800 & 10 \\
          \object{HE~1431$-$0245} &    14:33:54.2 &     -02:58:33  &   16.59  &   15.22         & 13.57  &   12.99  & 800 & 11 \\
          \object{HE~2138$-$1616} &    21:41:16.6  &    -16:02:40 & 14.81$^A$ &   13.68$^A$     & 11.90  &   11.32  & 350 & 9 \\
          \object{HE~2141$-$1441} &    21:44:25.7  &    -14:27:33  & 14.53$^A$ &  13.37$^A$     & 11.40  &   10.73  & 280 & 8 \\
          \object{HE~2144$-$1832} &    21:46:54.7  &    -18:18:15  &   12.65  &   11.14         &  8.77  &    7.96  & 30  & 6 \\
          \object{HE~2153$-$2323} &    21:56:37.6  &    -23:09:25  &   16.21  &   14.67         & 12.53  &   11.74  & 700 & 8 \\
          \object{HE~2155$-$2043} &    21:58:42.2  &    -20:29:15 &    13.92  &   13.16         & 11.57  &   11.02  & 300 & 40\\
          \object{HE~2235$-$5058} &    22:38:07.9  &    -50:42:42 &    13.83  &   12.92         & 11.43  &   10.89  & 200 & 10\\
          \object{HE~2250$-$4229} &    22:53:39.6  &    -42:13:03 & 12.66$^A$ &   11.91$^A$     & 10.40  &    9.87  & 100 & 30 \\
          \object{HE~2310$-$4523} &    23:13:00.0  &    -45:07:06 & 12.12$^A$ &   11.21$^A$     &  9.44  &    8.81  & 40  & 30\\
          \object{HE~2319$-$5228} &    23:21:58.1  &    -52:11:43 & 14.15$^A$ &   13.25$^A$     & 11.55  &   10.96  & 450 & 20  \\
          \object{HE~2357$-$2718} &    23:59:58.1  &    -27:01:37 & 14.31$^A$ &   13.08$^A$     & 11.09  &   10.42  & 170 & 6 \\
          \object{HE~2358$-$4640} &    00:00:50.9  &    -46:23:31 & 14.39$^A$ &   13.67$^A$     & 12.15  &   11.64  &  500 & 22 \\
  \hline
\end{tabular}
\tablefoot{Remarks: SB = single-lined spectroscopic binary. \\
$^A$ = APASS \citep{Henden2015}; $^a$ = average of APASS and values in 
Beers et al. (2007). }\\
\end{minipage}
\end{table*}


\section {Observations and data reduction}

The X-shooter spectrograph is described by \cite{Vernet2011}. It has
three arms: UV (300-550\,nm), visual (550-1000\,nm) and near-IR
(1000-2500\,nm). Our observations were obtained in March and/or August
2010. We used slits of 1\arcsec, 0.9\arcsec and 0.9\arcsec in the three
arms, yielding resolving powers $R$ = 4350, 7450, and 5300 in the UV,
visual, and near-IR spectral regions, respectively. The raw echelle
spectra were reduced with the X-shooter pipeline\footnote{Using the ESO
common pipeline libraries version 5.1.0.}, which is an automated routine
that performs the necessary data-reduction steps, including merging of
the orders. The one-dimensional output spectra were extracted, shifted
to rest wavelength using cross-correlation with an accuracy of
$\pm1$ km~s$^{-1}$, and the continua were normalised in IRAF by dividing 
the spectra with a fitted pseudo-continuum using cubic splines or Legendre
polynomials. 

\section{Stellar parameters and abundances}

The determination of stellar parameters and derivation of stellar
abundances were performed with the 1D LTE spectrum synthesis
code MOOG \citep[][version 2014]{Sneden1973}, using 1D interpolated
Kurucz NEW-ODF, ATLAS9 model atmospheres \citep{Castelli2003}, calculated
with the interpolation code described by \citet{Allende2004}.

\subsection{Stellar atmospheric-parameter estimates}

Extracting stellar atmospheric parameters from the spectra of stars that
are dominated by dense forests of molecular bands and absorption lines,
in particular for cool stars, is a challenging exercise. Most of the Fe
lines normally used for determining [Fe/H] or microturbulence are either
saturated or, at our resolving power ($R \sim 4350-7450$), so heavily blended
that we are left with just a handful of usable Fe lines. This is too
sparse to use the classical excitation and ionisation equilibrium
techniques for atmospheric-parameter estimates, especially since
the few useful Fe lines are limited to strong, low excitation-potential
lines. 

An obvious alternative tracer of [Fe/H] is the Ca triplet at $\sim$855
nm. In our case the triplet lines are of dubious value, as the nearby
TiO band is strong in most of our targets, interfering with the
continuum placement in this region. Moreover, metallicities derived from
the calibrations for red giant-branch stars by \cite{Cole2004} or for RR
Lyrae by \cite{Wallerstein2012} are 0.2 -- 1\,dex lower than indicated
by the few Fe lines we were able to measure. These apparently spuriously
low metallicities resulted in poor spectral fits when synthesizing the
stellar spectra using such metal-poor atmosphere models. 

After much experimentation, we concluded that the best option was either
to rebin the optical spectra so that they could be analyzed with
techniques developed for lower resolution ($R \sim 2000$) spectra, such
as those from SDSS, or else rely on the few useful Fe lines, as
described below.   

We estimated T$_{\rm eff}$, log $g$, and [Fe/H] using the n-SSPP, a
modified version of the SEGUE Stellar Parameter Pipeline
\citep[SSPP, see e.g.,][for a detailed description of the methods
employed]{Lee2008,Lee2013}. The n-SSPP uses low-resolution optical
spectra (typically covering the wavelength range 380-550\,nm),
which we have for our programme stars, and photometric information ($V_0$,
$(B-V)_0$, $J_0$, and $(J-K)_0$, corrected for extinction and reddening
based on the \citealt{Schlegel1998} dust maps), to determine
first-pass estimates for each parameter -- the [Fe/H] estimates
being particularly important. Typical internal errors for the
atmospheric parameters adopted by the n-SSPP are: 125~K for T$_{\rm
eff}$, 0.25~dex for log $g$, 0.20~dex for [Fe/H], and 0.25~dex for
[C/Fe]. External errors may well be larger, depending on the wavelength
coverage and S/N ratio of the input spectra \citep[see][for further
details]{Beers2014}. However, the n-SSPP pipeline fails for some stars
in our sample (in particular those with very low S/N or T$_{\rm eff} <
4500$~K, outside of the optimal range for the n-SSPP). In such cases, we
are forced to derive the stellar parameters in other ways.

The temperatures were finally determined using $V-K$ and the
IRFM-based calibrations of \citet{Alonso1999}. The E(B-V) were
downloaded from the IRSA webpage using the stellar IDs; we adopted
the mean E(B-V) values from \citet{Schlegel1998}. In the absence of
usable trigonometric parallaxes for our stars, we applied the log $g-$T$_{\rm eff}$
scaling relation from \citet{Barklem2005}. However, this relation yielded
unrealistically low gravities for some of the stars; we therefore
used isochrones, which in most cases agreed with the scaling 
relation to within 0.2\,dex.

We have employed BaSTI \citep[e.g.,][]{BaSTI2013} and Padova
\citep{Girardi2000} isochrones with the appropriate metallicities and
C-N-O-enhancements to calculate surface gravities, adopting the
photometric temperatures and [Fe/H] from the n-SSPP pipeline. Since the
BaSTI isochrones resulted in somewhat larger values of log $g$ for nine
stars than expected from the n-SSPP pipeline, we adopted Padova
isochrones for metallicities in the range $Z = 0.01 - 0.004$. A lower Z
in the adopted isochrone leads to a lower gravity. The initial
[Fe/H] estimates from the n-SSPP pipeline were adopted, and only adjusted
if the strong Fe lines close to Sr and Ba (e.g., 406.3 and 407.1\,nm)
indicated a much lower or higher metallicity. In these instances we carried
out a spectrum synthesis to fit these Fe lines and updated the [Fe/H]
values accordingly. Since these Fe lines are strong, this approach
may introduce a bias in our [Fe/H], but these lines remain detectable in our low-S/N, moderate-resolution spectra.

In summary, the stellar parameters were determined using
photometry, Padova isochrones, and an [Fe/H] estimate based on a
combination of the n-SSPP output and by fitting synthetic spectra to a
few strong Fe lines near the Sr and Ba lines. The microturbulence was
estimated by applying the Gaia-ESO scaling relation between temperature,
gravity, metallicity, and microturbulence (Bergemann et al. in prep.). 
This resulted in values
typically between 1.5 and 2.0\, km\,s$^{-1}$ for most stars. The
adopted stellar parameters are thus more uncertain than those derived
from high-resolution spectra, and these uncertainties will propagate
into the derived stellar abundances (see Sect. \ref{error}). The final
atmospheric parameters are listed in Table~\ref{table:param}. 

\begin{table}
\centering
\caption{Atmospheric parameters for the programme stars; 
`$^u$' indicates uncertain values.}
\label{table:param} 
\begin{tabular}{llllc}
\hline
Star & T$_{eff}$  & log $g$ & [Fe/H] & $\xi$ \\
  & [K] & [cgs]  &  [dex] & [km~s$^{-1}$]\\
\hline
          \object{HE~0058$-$3449} &       5400 &   3.0  & $-2.1$ & 1.5 \\
          \object{HE~0206$-$1916} &       4800 &   1.4  & $-2.4$ & 2.0 \\
          \object{HE~0241$-$3512} &       4600 &   1.2  & $-1.8$ & 2.0\\
          \object{HE~0400$-$2030} &       5200 &   2.5  & $-2.2$ & 1.7\\  
          \object{HE~0408$-$1733} &       4600 &   1.4  & $-0.8^u$ & 1.7 \\  
          \object{HE~0414$-$0343} &       4700 &   1.3  & $-2.5$ & 2.0 \\
          \object{HE~0430$-$1609} &       4700 &   1.3  & $-2.3$ & 2.0 \\
          \object{HE~0430$-$4901} &       5500 &   3.3  & $-3.1$ & 1.5 \\
          \object{HE~0440$-$3426} &       4700 &   1.3  & $-2.6$ & 2.0\\
          \object{HE~0448$-$4806} &       5800 &   3.4  & $-2.4$ & 1.5\\
          \object{HE~0516$-$2515} &   $4400^u$ &   0.7  & $-2.5^u$ & 2.2\\
          \object{HE~1238$-$0836} &   4100$^u$ &$0.1^u$ & $-1.9$ & 2.3\\
          \object{HE~1315$-$2035} & $5100^u$& $2.2^u$ & $-2.5^u$ & 1.8\\
          \object{HE~1418$+$0150} &       4200 &   0.6  & $-1.3^u$ & 2.2\\
          \object{HE~1430$-$0919} &       4900 &   1.6  & $-2.5$ & 1.9\\
          \object{HE~1431$-$0245} &       5200 &   2.3  & $-2.5^u$ & 1.8\\
          \object{HE~2138$-$1616} &       4900 &   1.9  & $-0.5$ & 1.6\\
          \object{HE~2141$-$1441} &       4600 &   1.4  & $-0.6$ & 1.6\\
          \object{HE~2144$-$1832} &       4200 &   0.6  & $-1.7$ & 2.2\\
          \object{HE~2153$-$2323} &       4300 &   0.6  & $-2.4^u$ & 2.3\\
          \object{HE~2155$-$2043} &       5200 &   2.4  & $-3.0$ & 1.5\\
          \object{HE~2235$-$5058} &       5200 &   2.5  & $-2.7$ & 1.5\\
          \object{HE~2250$-$4229} &       5200 &   2.4  & $-2.7$ & 1.7\\
          \object{HE~2310$-$4523} &       4700 &   1.4  & $-2.5$ & 2.0\\
          \object{HE~2319$-$5228} &       4900 &   1.6  & $-2.6$ & 2.0\\
          \object{HE~2357$-$2718} &       4500 &   1.3  & $-0.5$ & 1.7\\
          \object{HE~2358$-$4640} &       5100 &   2.4  & $-1.7$ & 1.6\\
  \hline
\end{tabular}
\end{table}

\subsection*{Line lists}

Before synthetic spectra can be generated, line lists for all relevant
features must be assembled. The line list for the molecular bands such
as NH, CH, C$_2$, and CN are taken from \citet{Masseron2014} and T.
Masseron (2015, priv. comm.). The adopted dissociation energies are
3.47\,eV (CH), 3.42\,eV (NH), 6.24\,eV (C$_2$), and 7.7\,eV (CN). The
line list covering the atomic lines (Sr and Ba) are from
\citet{Sneden2014}, with the Sr hyperfine structure from
\citet{Bergemann2012} and \citet{Hansen2013}, and the Ba hyperfine structure (HFS) from
\citet{Gallagher2012}. Since most of these stars are of higher
metallicities or enriched by an $s$-process, we used the total
Solar System isotopic-abundance ratios for the heavy elements.

\begin{figure}[!ht]
\begin{center}
\includegraphics[width=0.45\textwidth]{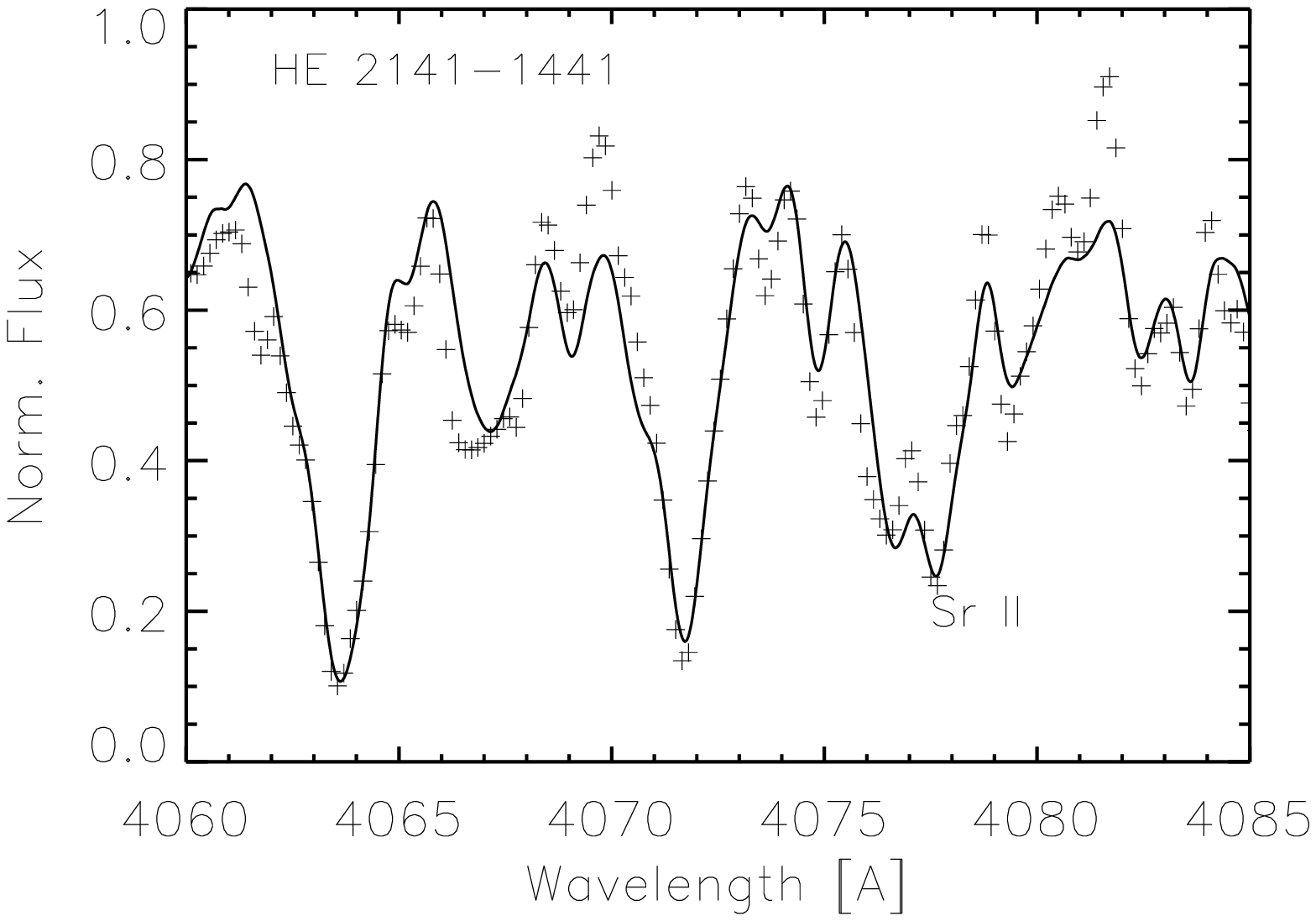}
\includegraphics[width=0.45\textwidth]{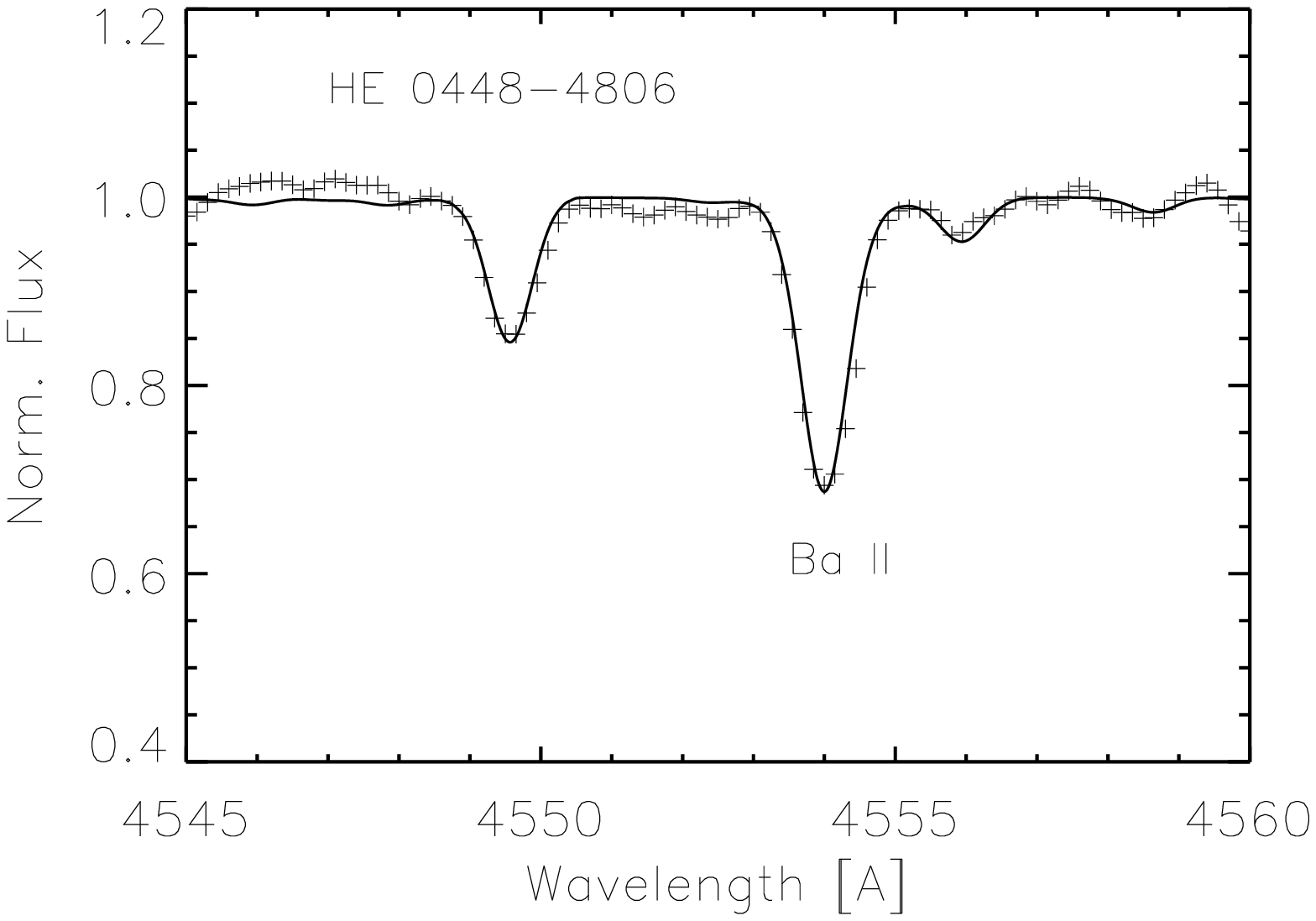}
\caption{Synthetic spectrum fits to the Sr II 407.7\,nm 
(upper panel) and Ba II 455.4\,nm lines (lower panel). 
The observations of \object{HE~2141$-$1441} (relatively metal rich, $s$-element poor) and
\object{HE~0448$-$4806} (relatively metal poor, $s$-element rich) are shown as
black plus signs. }
\label{SrBaspec}
\end{center}
\end{figure}

\subsection{Derivation of molecular and atomic stellar abundances}

With the ATLAS9 model atmospheres, the above line list, and the stellar
parameters in Table \ref{table:param}, the synthetic spectra for
each star were computed with MOOG and convolved with a Gaussian
representing the X-shooter resolution at the appropriate wavelength.

Synthetic spectra of the CH $G$-band (430\,nm) and C$_2$ Swan band
(516\,nm) were fit to the observations to derive the [C/Fe] abundance
ratios of our stars, using $\chi^2$ minimization for matching the
synthetic and the observed spectra across a selected sensitive spectral
interval. Nitrogen abundances were estimated in the same way, by fitting
synthetic spectra to the NH band around 336\,nm and CN around 389\,nm
and 421\,nm, respectively. The listed abundances are average values of
these measurements. The stars with low C or N rely only on the strongest
CH or NH band, respectively.

The isotopic carbon ratios were derived by measuring molecular $^{13}$CH
features in the range 421--423.5\,nm.

Originally, we intended to obtain estimates of O from the CO bands in
the near-IR, but the quality of the spectra obtained proved
insufficient. We hope to obtain improved spectra in the future, from
which another attempt to extract O abundances will be made.

The neutron-capture elements Sr and Ba are of particular interest for
CEMP stars, as they enable us to distinguish between the
CEMP-$s$ and CEMP-no sub-classes. Both elements are detectable in the
X-shooter spectra, due to their strong resonance lines
\citep{Hansen2013}. Their abundances were derived via line-by-line
spectrum synthesis with line lists including all relevant molecular and
atomic lines, and taking into account the HFS of both the Sr and Ba
lines. The Sr II line at 407.7\,nm is the only useful line of Sr, as all
other lines are either too weak or are obliterated by strong blends or
molecular bands. However, this line is very strong and tends to
saturate. 

For Ba, we used the two lines at 455.4\,nm and 585.3\,nm, since
they generally yield abundances in good agreement with each other. The
455.4\,nm line is the strongest, and is usually detected even in the
most metal-poor stars. The 585.3\,nm line is weaker, and for that reason
possibly more reliable when detectable. The 493.4\,nm line is too
blended and/or weak to yield reliable abundances, and it was only used
to support the upper limit values estimated from the 455.4\,nm line when
the 585.3\,nm line was too weak or blended. Line blends were taken into
account by scaling the blends to the overall [Fe/H], $\alpha$-, $r$- and
$s$-process levels when deriving Sr and Ba abundances. Figure
\ref{SrBaspec} shows our spectrum synthesis around the Sr and Ba lines
for two programme stars.

\subsection{Error propagation\label{error}}

The uncertainties listed in Table~\ref{tab:results} indicate how
well the stellar parameters could be determined, and in turn, how well
the abundances could be derived. Several of the stars show signs of
variability or are known binaries; in most cases these could not be fit as 
well as the constant VMP stars. The listed uncertainties on the temperature
estimates take into account the variability (which could change the 
magnitudes and thereby the colour of the star) as well as uncertainties 
on the de-reddening. These uncertainties are folded into those on
the gravities, where a good agreement between Padova and BaSTI
isochrones, as well as an agreement between these and the n-SSPP and the
scaled gravities are also taken into account (i.e., the standard
deviation around the mean log $g$ is considered as well). The
uncertainties on the metallicity are estimated from a line-by-line
synthesis of the clearly detectable Fe lines in the vicinity of Sr and
Ba. The uncertainties have been rounded off in the same way as the
stellar parameters.

At the resolving power in the blue arm of X-shooter ($R \sim 4350$) all
the features of interest are blended to some extent. In
particular, features within $\sim \pm$0.1\,nm, are blended into the two
Sr and Ba lines. These blends affect the Sr line (407.7\,nm) the
most, and we compensated for the blending by La and Dy lines by setting
their abundances to that derived from Ba, and then adjusted this value by up
to 0.3\,dex to obtain a better fit of the Sr line.

A minimum abundance error of 0.1 dex due to blends, continuum
placement, line list uncertainties, and overall spectrum quality
 is included in the total error budget, rising to 0.2 dex for the
bluest regions of the lowest-quality spectra. For poor-quality
spectra, the largest uncertainties are due to the stellar parameters
(T/log$g$/[Fe/H]/$\xi$: $\pm300$\,K/ $\pm0.3$\,dex/ $\pm0.3$\,dex/
$\pm0.3$\,km s$^{-1}$). In the following we describe our detailed
analysis of the maximum abundance uncertainties caused by the stellar
parameters for a spectrum of low quality (see Tables~\ref{table:param}
and \ref{uncert}).

\begin{table}
\centering
\caption{Abundance sensitivity to stellar parameters for \object{HE~0400$-$2030}.}
\label{uncert} 
\begin{tabular}{lccc}
\hline \hline

Line     & Sr 407.7 nm  &  Ba 455.4 nm  & Ba 585.3 nm \\
Abundance$:$ &  [dex] &  [dex] &  [dex]\\
\hline
Parameter: \\
T$\pm$300~K     &  0.32     &  0.35      &   0.3    \\
Log $g\pm$0.3\,dex  &  0.03     &  0.05       &   0.05     \\
$[$Fe/H$]\pm$0.3\,dex &  0.26      &  0.15      &   0.25     \\
$\xi \pm$ 0.3\,km~s$^{-1}$ &  0.03     &  0.02       &   0.15     \\
\hline
Max. uncert.$:$& 0.41 & 0.38 & 0.42 \\
\hline 
Mean uncert.$:$& 0.34 & 0.3 & 0.3 \\ 
\hline \hline
\end{tabular}
\end{table}

A minimum uncertainty was estimated using the best spectra and the
best-constrained stellar atmospheric-parameter estimates, and is on the order of
T/log$g$/[Fe/H]/$\xi:$ 100~K/ 0.2\,dex/ 0.1\,dex/ 0.15 km s$^{-1}$,
similar to the uncertainty seen in high-resolution studies and
corresponding to our best spectra with a S/N ratio $>30$. In this study,
most of the stars actually have spectra with S/N $\sim10$ at 400\,nm,
leading to larger errors. From the minimum and maximum uncertainties we
calculated a mean representing the general abundance uncertainty for Sr
and Ba.

As seen in Table~\ref{uncert}, the temperature has the largest impact on
the derived abundances. However, Sr and Ba are ionised, so gravity also
influences the line strength, but less visibly due to the blends in the
line wings. Microturbulence has the least impact on the Sr and 455.4\,nm
Ba lines, while the 585.3\,nm line of Ba is clearly affected by this
parameter. The difference in stellar-parameter sensitivity, combined
with line blends and varying spectral resolution, explains the small
difference in Ba abundances derived from the 455.4\,nm line in the blue
region ([Ba/Fe] = 1.95) and the 585.3\,nm Ba line ([Ba/Fe] $\sim 1.7$) in
the visual region. However, all the Sr and Ba lines agree on the
presence or absence of any $s$-process enhancement, so a star remains
enhanced whether or not we consider the blue or visual region of the
spectrum.

Finally, we adopted a minimum error of 0.2\,dex for the C and N
abundances, dominated by the uncertainties in the atmospheric-parameter
estimates.

\section{Results}

 The derived abundances are listed in Table~\ref{tab:results}. 
The CEMP classification of \citet{Beers2005}, but adopting 
a lower limit of [C/Fe] $> 0.7$ as the criterion for CEMP
stars \citep{Aoki2007}, leads to 17 CEMP-$s$ stars, three CEMP-no stars, 
and seven carbon-normal metal-poor stars\footnote{Throughout the paper we refer to C-normal very metal-poor stars as VMP stars, which are to be distinguished from the C-enhanced CEMP stars.} in the total sample.
Some of the stars are difficult to fit owing to low S/N, binarity, or
pulsations; these are identified by a '$^u$' in the table to indicate that the
results are more uncertain. The poorer fit to these stars is characterised 
from a larger standard deviation between the observations and the synthetic 
MOOG spectra in the range 406 -- 409\,nm.

\begin{table*}
\centering
\hspace{-1cm}
\setlength{\tabcolsep}{4pt} 
\caption{Final stellar parameters, C, N, Sr, and Ba abundances, $^{12}$/$^{13}$C isotopic ratios, and CEMP classes.}
\label{tab:results} 
\begin{tabular}{lcccrrrrccc}
\hline
\hline
 Star 	& T$_{eff}$ & log $g$ & [Fe/H] &  [C/Fe] & [N/Fe] & [Sr/Fe] & [Ba/Fe] & $^{12}$/$^{13}$C &CEMP & Note\\   
 & [K] & [cgs] &  &   & &  & &  & class & \\   
\hline
\object{HE~0058$-$3449} &       $5400\pm200$ &   $3.0\pm 0.1$   & $-2.1\pm0.2$  &  $1.0\pm0.2$    &  $1.1\pm0.2$    &	 $1.3^u\pm0.3$    &  $1.8\pm0.2$   &	9  &  $s$	&     \\
\object{HE~0206$-$1916} &       $4800\pm200$ &   $1.4\pm 0.1$   & $-2.4\pm0.2$  &  $1.9\pm0.2$    &  $1.1\pm0.3$    &	 $0.9\pm0.2$	  &  $1.3\pm0.3$   &   24  &  $s$	&	\\
\object{HE~0241$-$3512} &       $4600\pm100$ &   $1.2\pm 0.1$   & $-1.8\pm0.2$  &  $1.3\pm0.2$    &  $0.3\pm0.2$    &	 $1.6\pm0.2$	  &  $2.0\pm0.2$   &	9  &  ($s$)	&	\\
\object{HE~0400$-$2030} &       $5200\pm200$ &   $2.5\pm 0.1$   & $-2.2\pm0.1$  &  $1.1\pm0.2$    &  $2.5\pm0.2$    &	 $1.7\pm0.2$	  &  $2.0\pm0.2$   &	4  &  $s$	& NEMP  \\
\object{HE~0408$-$1733} &       $4600\pm100$ &   $1.4\pm 0.3$   & $-0.8\pm0.3$  &  $0.3\pm0.2$    &  $-0.6\pm0.3$   &    $0.6\pm0.2$      &  $0.2\pm0.2$   &   --  &   MP	 &	 \\
\object{HE~0414$-$0343} &       $4700\pm100$ &   $1.3\pm 0.1$   & $-2.5\pm0.2$  &  $1.3\pm0.2$    &  $1.6\pm0.2$    &    $1.6\pm0.2$      &  $1.9\pm0.2$   &	9  &  $s$	 &  *,b (1)	  \\
\object{HE~0430$-$1609} &       $4700\pm300$ &   $1.3\pm 0.2$   & $-2.3\pm0.2$  &  $1.8\pm0.2$    &  $0.4\pm0.3$    &    $1.6\pm0.3$      &  $ < 2.4$     &   --  &  $s$	  &  u,b (2)    \\
\object{HE~0430$-$4901} &       $5500\pm100$ &   $3.3\pm 0.3$   & $-3.1\pm0.2$  &  $1.0\pm0.3$    &  $1.1\pm0.3$    &    $1.0\pm0.2$      &  $ < 1.5$     &   --  &  $s$	  &	  \\
\object{HE~0440$-$3426} &       $4700\pm200$ &   $1.3\pm 0.2$   & $-2.6\pm0.3$  &  $1.6\pm0.2$    &  $1.1\pm0.2$    &    $0.9\pm0.2$      &  $1.2\pm0.2$   &    9  &  $s$	  &	  \\
\object{HE~0448$-$4806} &       $5800\pm200$ &   $3.4\pm 0.2$   & $-2.4\pm0.3$  &  $2.2\pm0.2$    &  $1.2\pm0.3$    &    $1.2\pm0.2$      &  $2.3\pm0.2$   &    5  &  $s$	  &	  \\
\object{HE~0516$-$2515} &       $4400\pm100$ &   $0.7\pm 0.3$   & $-2.5\pm0.3$  &  $> 0.8    $    &   --            &    $0.5^u\pm0.3$    &  $0.5\pm0.3$   &   --  & ($s$/no) & u,*      \\
\object{HE~1238$-$0836} &       $4100\pm300$ &   $0.1\pm 0.3$   & $-1.9\pm0.3$  &  $> 0.2    $    &   --            &    $0.3^u\pm0.3$    &  $ > -1.5$     &   --  &   MP       & u, R Sct (A)\\
\object{HE~1315$-$2035} &       $5100\pm300$ &   $2.2\pm 0.3$   & $-2.5\pm0.2$  &  $2.5\pm0.3$    &  $1.1\pm0.2$    &    $1.6\pm0.3$      &  $ < 2.6$      &   19  &  $s$      & u      \\
\object{HE~1418$+$0150} &       $4200\pm100$ &   $0.6\pm 0.3$   & $-2.3\pm0.3$  &  $> +0.1    $    &   --           &    $1.0\pm0.3$      &  $ > 1.0$       &   --  &   MP(s)      & u       \\
\object{HE~1430$-$0919} &       $4900\pm200$ &   $1.6\pm 0.2$   & $-2.5\pm0.1$  &  $2.2\pm0.2$    &  $0.8\pm0.3$    &    $1.0\pm0.2$      &  $1.5\pm0.3$   &   32  &  $s$	  &	  \\
\object{HE~1431$-$0245} &       $5200\pm200$ &   $2.3\pm 0.2$   & $-2.5\pm0.3$  &  $2.4\pm0.2$    &  $1.5\pm0.3$    &    $1.6\pm0.2$      &  $1.9\pm0.2$   &   49  &  $s$	  &	   \\
\object{HE~2138$-$1616} &       $4900\pm200$ &   $1.9\pm 0.3$   & $-0.5\pm0.3$  &  $0.2\pm0.3$    &  $0.0\pm0.3$    &    $0.7\pm0.3$      &  $-0.4\pm0.2$  &   --  &   MP	  & u	   \\
\object{HE~2141$-$1441} &       $4600\pm100$ &   $1.4\pm 0.3$   & $-0.6\pm0.3$  &  $0.2\pm0.2$    &  $0.2\pm0.3$    &    $0.5^u\pm0.3$    &  $0.1\pm0.3$   &   --  &   MP	  &	  \\
\object{HE~2144$-$1832} &       $4200\pm100$ &   $0.6\pm 0.3$   & $-1.7\pm0.3$  &  $0.8\pm0.2$    &  $0.6\pm0.2$    &    $1.5\pm0.2$      &  $1.3\pm0.2$   &   --  &  ($s$)    & CH star (A) \\
\object{HE~2153$-$2323} &       $4300\pm100$ &   $0.6\pm 0.2$   & $-2.4\pm0.3$  &  $1.6\pm0.2$    &  $0.7\pm0.3$    &    $1.2\pm0.2$      &  $1.1\pm0.2$   &   12  &   $s$	   &	   \\
\object{HE~2155$-$2043} &       $5200\pm100$ &   $2.4\pm 0.2$   & $-3.0\pm0.2$  &  $0.7\pm0.2$    &  $0.0\pm0.2$    &    $0.2\pm0.2$      &  --	           &   --  &   no	   & *    \\
\object{HE~2235$-$5058} &       $5200\pm100$ &   $2.5\pm 0.2$   & $-2.7\pm0.2$  &  $1.0\pm0.2$    &  $1.0\pm0.2$    &    $1.7\pm0.2$      &  $2.4\pm0.2$   &   32  &   $s$	   &	  \\
\object{HE~2250$-$4229} &       $5200\pm100$ &   $2.4\pm 0.2$   & $-2.7\pm0.3$  &  $0.9\pm0.3$    &  $0.4\pm0.3$    &    $-0.3\pm0.2$     &  $-0.3\pm0.2$  &   --  &   no	   &	   \\
\object{HE~2310$-$4523} &       $4700\pm200$ &   $1.4\pm 0.2$   & $-2.5\pm0.2$  &  $0.2\pm0.3$    &  $0.0\pm0.3$    &    $0.0^u\pm0.3$    &  $-0.4\pm0.2$  &   --  &   VMP	   &	   \\
\object{HE~2319$-$5228} &       $4900\pm100$ &   $1.6\pm 0.2$   & $-2.6\pm0.2$  &  $1.7\pm0.2$    &  $2.5\pm0.2$    &    $ < -3 $	  &  $ < -3$       &    5  &   no	   &	   \\
\object{HE~2357$-$2718} &       $4500\pm100$ &   $1.3\pm 0.3$   & $-0.5\pm0.1$  &  $0.4\pm0.3$    &  $-0.1\pm0.3$   &    $0.3\pm0.2$      &  $-0.3\pm0.2$  &   --  &   MP	   &	   \\
\object{HE~2358$-$4640} &       $5100\pm200$ &   $2.4\pm 0.1$   & $-1.7\pm0.2$  &  $0.2\pm0.3$    &  $-0.1\pm0.3$   &    $0.1\pm0.2$      &  $-0.3\pm0.3$  &   --  &   MP	   & *      \\
\noalign{\smallskip}
\hline
\noalign{\smallskip}
\hline
\end{tabular}
\tablefoot{\tablefoottext{*}{Stellar parameters agree with values from n-SSPP pipeline.}\tablefoottext{b}{Spectroscopic binary.} \tablefoottext{u}{Uncertain values.}\tablefoottext{A}{Possibly an AGB star.}\tablefoottext{1}{Binary according to Hollek et al. 2015 (accepted).}\tablefoottext{2}{Binary according to Hansen et al. 2015c, (submitted).}}\\
\end{table*}

\subsection {C and N}

In addition to the elemental abundances of C and N, we estimated 
carbon isotope ratios, $^{12}$C/$^{13}$C, for 13 of our programme stars 
from $^{13}$CH lines in the range 421 -- 423.5\,nm; one CEMP-no and 12 CEMP-$s$ stars (see Table~\ref{tab:results}).

All of the stars labelled as CEMP-$s$ have [C/Fe]
$> 0.7$ and [Ba/Fe] $> 1.0$, which rules out their association with
the CEMP-no class. However, three stars exhibit weak Sr absorption lines
and almost no Ba features. These are so faint that we can only 
provide upper limits ([Sr,Ba/Fe] $< -3$) in one case. These three CEMP-no stars are
\object{HE~2155$-$2043}, \object{HE~2250$-$4229}, and \object{HE~2319$-$5228},
discussed in detail in Sect.~\ref{sec:CEMPno}. We also identify one very
nitrogen-enhanced metal-poor (NEMP) star, \object{HE~0400$-$2030}. 

\begin{figure}[!ht]
\begin{center}
\includegraphics[width=0.49\textwidth]{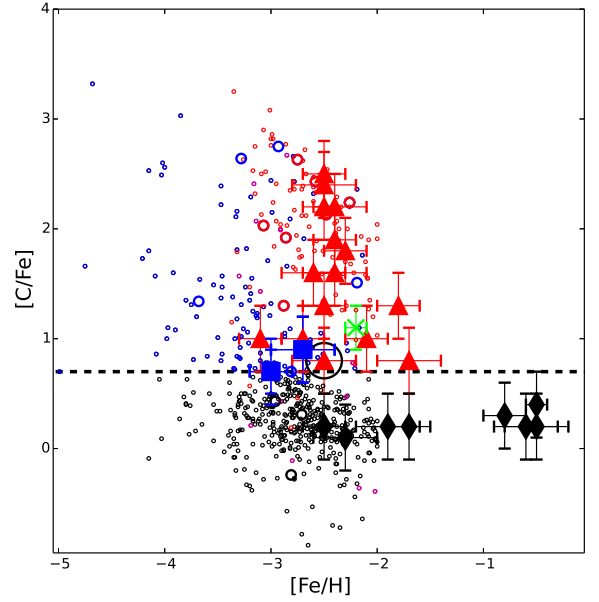}
\caption{ [C/Fe] vs. [Fe/H] for our sample, compared to
literature values: \citet{THansen2015} (large, open circles) and
\citet{Placco2014b} and references therein (small, open circles). 
CEMP-$s$ stars are shown in red,  CEMP-no stars in blue, the NEMP star in green, and carbon-normal 
MP stars in black.
The dashed line at [C/Fe] = 0.7 separates the VMP and CEMP stars.}
\label{comptolit}
\end{center}
\end{figure}

\begin{figure}
\centering
\includegraphics[width=0.49\textwidth]{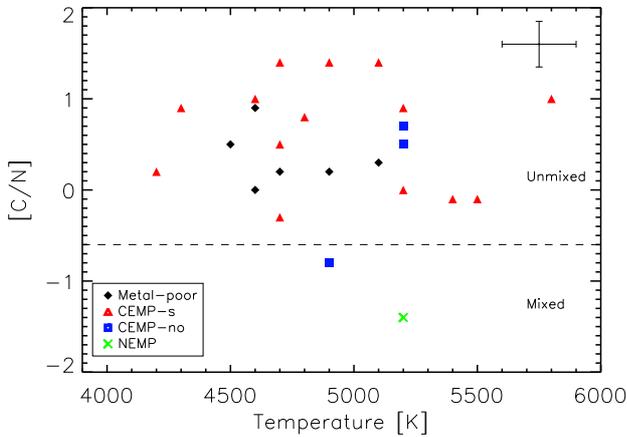}
\caption{[C/N] ratios determined from our X-shooter spectra. C-normal 
MP stars are shown as black diamonds and CEMP-$s$ and CEMP-no as red triangles and blue 
squares, respectively, while the green cross shows the NEMP star. 
The dashed line indicates the limit below which 
CNO-cycled material has been mixed to the surface.}
\label{CNmixing}
\end{figure}

All our programme stars have been checked against the data of
\citet{Placco2014b} and \citet{THansen2015}, and the comparison samples
used therein (Fig.~\ref{comptolit}). Figure~\ref{CNmixing} shows the
[C/N] abundance ratio for our programme stars. Two stars
(\object{HE~0400$-$2030} and \object{HE~2319$-$5228}; discussed in more detail in
Sect.~\ref{sec:CEMPno}) are found to have mixed CNO-cycled material to
their surfaces, converting some C into N \citep{Spite2005,Spite2006},
while C is normal or enhanced in the rest of the stars. We note that
since most of our stars have relatively high gravities (and are
subgiants), we do not expect that mixing would alter the surface
composition of, e.g., C, and in most cases the C
corrections\footnote{Correction for stellar evolution effects that convert 
C into N.} from \citet{Placco2014b} are negligible (on the order of 0.0
- 0.03\,dex). Even for such low-gravity stars as \object{HE~2144$-$1832}, the
correction obtained is only $\sim0.13$\,dex. Therefore, we have not applied
these corrections to any of our abundances. 

\begin{figure*}[!ht]
\begin{center}
\includegraphics[width=0.99\textwidth]{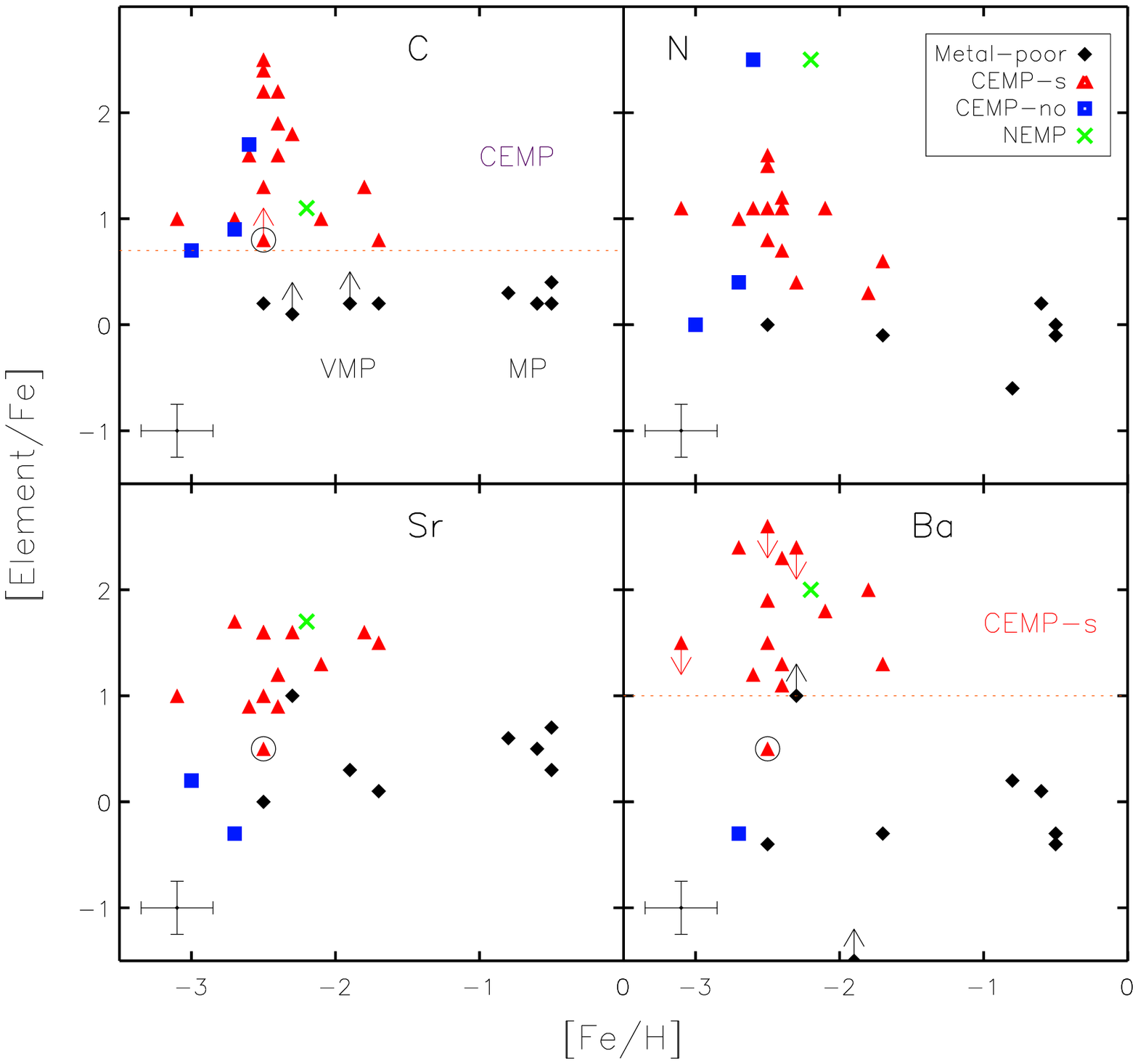}
\caption{[C, N, Sr, Ba/Fe] ratios vs. [Fe/H] for our programme
stars. Chemically normal ([C/Fe]$< 0.7$) stars are shown in black; 
 red triangles represent CEMP-$s$ stars, and blue squares are CEMP-no stars, and 
the green 'x'  symbol is the single NEMP star, respectively. The circle marks the 
uncertain CEMP-no/$-s$ star \object{HE~0516$-$2515}. Dotted lines at [C/Fe] $= 0.7$ and [Ba/Fe] $= 1.0$ separate the VMP from the CEMP(-$s$) stars. \label{Multifig}} 
\end{center}
\end{figure*}

A summary of our abundance results for C and N is shown in the two top 
panels of Fig.~\ref{Multifig} as functions of [Fe/H].

\subsection {Sr and Ba}

The Sr and Ba abundances help to distinguish the CEMP-$s$ stars, which 
have presumably received C-enriched matter transferred from a former AGB
companion \citep{Lucatello2005}, from the chemically normal VMP and
CEMP-no stars. In contrast, the normal binary frequency of the CEMP-no stars 
(\citet{Starkenburg2014,Bonifacio2015} and Hansen et al. 2015b, subm.) suggests that their abundances
reflect the composition of the gas from which they were born.

The derived abundances of Sr and Ba are shown as functions of [Fe/H] 
in the two lower panels of Fig.~\ref{Multifig}, 
in which the metal-poor stars 
[C/Fe]$ < 0.7$) are identified by black symbols. At [Fe/H] $<
-1.5$, clear divisions of the Sr and Ba abundances into two branches are
seen. Here, the span in Sr and Ba abundances at the lowest [Fe/H] values
amounts to $\sim 3$\,dex \citep[as also found in][]{Hansen2012},
corresponding to the difference between the chemically normal and
CEMP-no stars on the one hand, and the highly enhanced CEMP-$s$ stars on
the other\footnote{Mass transfer from a binary companion seems to be the only efficient way to change the surface composition of the CEMP-$s$ stars to the observed levels.}. 
Between the branches is the CEMP-$no/s$ star 
\citep[for a definition see][]{Sivarani2006} \object{HE~0516$-$2515}, with [Ba/Fe] $= 0.5$ and [Fe/H] = $-2.5$. Our X-shooter 
results for this star are uncertain, and higher-resolution observations 
are required to classify it 
with confidence. The blue data points on the low-Sr branch in
Fig.~\ref{Multifig} are the CEMP-no stars \object{HE~2155$-$2043} and
\object{HE~2250$-$4229}, while \object{HE~2319$-$5228} falls below this branch. 
Except for \object{HE~2319$-$5228}, the CEMP-no stars have relatively low
[C/Fe] compared to the CEMP-$s$ stars.

\subsection{Comparison to literature}\label{lit:RVsct}

We have several stars in common with previous studies. Our temperatures
generally agree within 150\,K with those presented in
\citet{Goswami2005}, \citet{Goswami2010}, and \citet{Kennedy2011}.
The light-element (C, N) abundances agree within 0.2\,dex,
 while larger differences (up to 0.4\,dex)
are found among the heavy-element abundances. This large difference
can be accounted for by differences in the adopted gravities and
metallicities, line lists, and continuum placement. The low resolution of the 
spectra classified in the study by \citet{Goswami2010} prevented them from
deriving accurate metallicities; they estimated these values by 
comparing the spectra of their high-latitude stars to those
with better-known parameters. 

Our metallicities also generally agree with
those presented in \citet{Kennedy2011} and \citet{Aoki2007}, within 0.3\,dex. The
latter study is based on a high-resolution spectral analysis, compared to
which we find fair agreement between most of our results except from
the temperatures (owing to differences in adopted E($B-V$) values). 

All abundances and other parameters derived for \object{HE~1238$-$0836} 
are uncertain. The spectra for this star are of low quality, and furthermore, 
they resemble those of variable RV Tau-type stars\footnote{Since the star is bright, it may
belong to the bright sub-class R Sct.} according to \citet{Goswami2010}. 
The pulsations in such a star can lead to large amplitude (and thus magnitude) changes depending 
on sub-class. Such variations would in turn lead to very different stellar 
parameters and abundances, depending on whether the star was observed 
in an expanding or contracting phase.

\citet{Goswami2010} listed temperatures in the range between 3500 and 4000\,K 
for \object{HE~1238$-$0836}. If the star were as cool as 3500\,K when observed 
with X-shooter, we would
expect to see very strong TiO bands in the red spectra (as in M dwarfs),
which we do not. Based on the line profiles (Balmer lines as well as Ba
lines) we believe that the star has not been observed at a quiet
phase, but we need higher-resolution observations during this phase to
improve our abundance measurements.

\section{Discussion}

The recent study of CEMP stars by \citet{Spite2013} showed that,
depending on the level of absolute carbon abundance, the stars split
into two bands (which they referred to as ``plateaus''), where the
strongly C-enhanced stars are typically CEMP-$s$ stars, while the
relatively less C-enriched stars are CEMP-no stars. Several studies have
confirmed the existence of the carbon bands, and have populated them with
more stars \citep[e.g., ][]{Bonifacio2015, THansen2015}. Here we
discuss the behaviour of the heavy elements, and how they associate with
these bands, in order to better constrain the astrophysical sites and
processes that enriched the different sub-classes of CEMP stars.
Finally, since our sample also contains carbon-normal VMP stars, we
comment on similarities and differences in the formation of VMP stars
vs. CEMP stars.

\begin{figure}
\centering
\includegraphics[width=0.49\textwidth]{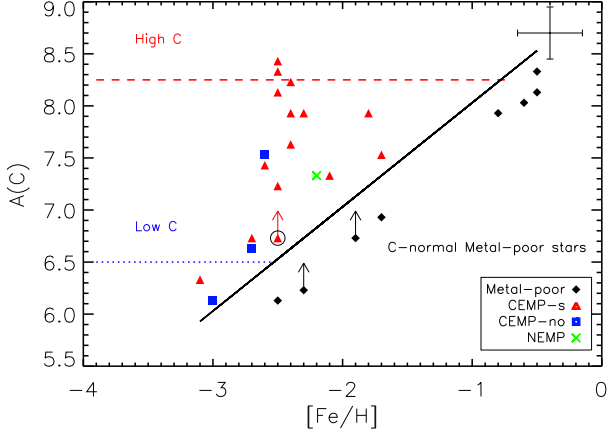}
\caption{ Absolute carbon abundances, $A(\rm C)=\log\epsilon\,(C)$, of our programme stars, as a function of metallicity, [Fe/H]. The high-C and low-C bands of \citet{Spite2013} are indicated by the
horizontal lines. The \citet{asplund2009} solar abundance of carbon, $\log\epsilon$ = 8.43, has been assumed. }
\label{plateau}
\end{figure}

Figure~\ref{plateau} shows the absolute C abundances for our programme
stars, as a function of metallicity, [Fe/H]. The VMP stars are
located below the [C/Fe] $= 0.7$ line.  Roughly half of the CEMP-$s$
stars lie close to the high-C band ($A\rm{(C)} = 8.25$), while two of
our CEMP-no stars are close to the low-C band ($A\rm{(C)} = 6.25$). The
other CEMP-no stars, the NEMP star, and a few CEMP-$s$
stars, fall between the two bands. We do not find any of our programme
CEMP-no stars located near the high-C band. Our results, as well as those from other larger studies indicate that there may exist a continuum of
absolute C abundances lying between the high-/low-C bands, blurring
their distinction.
\begin{figure}
\centering
\includegraphics[width=0.49\textwidth]{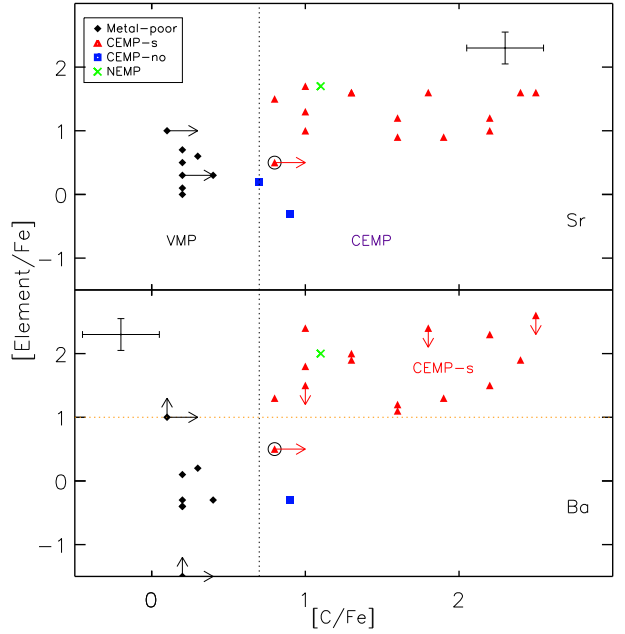}
\caption{[Sr/Fe] (top) and [Ba/Fe] (bottom) vs. [C/Fe]. The dashed line in the lower panel indicates the limit for $s$-process enhanced stars at [Ba/Fe] $>$ 1.0. The vertical line at [C/Fe]$=0.7$ separates VMP from CEMP stars.} 
\label{SrBaCFe}
\end{figure}

A discontinuity is seen in both panels of Fig.~\ref{SrBaCFe}, which 
illustrates the separation of CEMP-$s$ vs. CEMP-no and VMP 
stars. 
The large difference in neutron-capture abundance of 1-2\,dex found 
between C-normal VMP and CEMP-no stars and the CEMP-$s$ stars
explains some of the large star-to-star scatter found at low
metallicity. This scatter has been found in numerous
previous studies, and points towards differences in the formation sites
and/or processes involved, if these are assumed to be robust and produce
similar amounts of heavy elements in each event \citep[see, e.g.,
][]{Hansen2014b}.

\begin{figure}[h!]
\begin{center}
\includegraphics[width=0.33\textwidth]{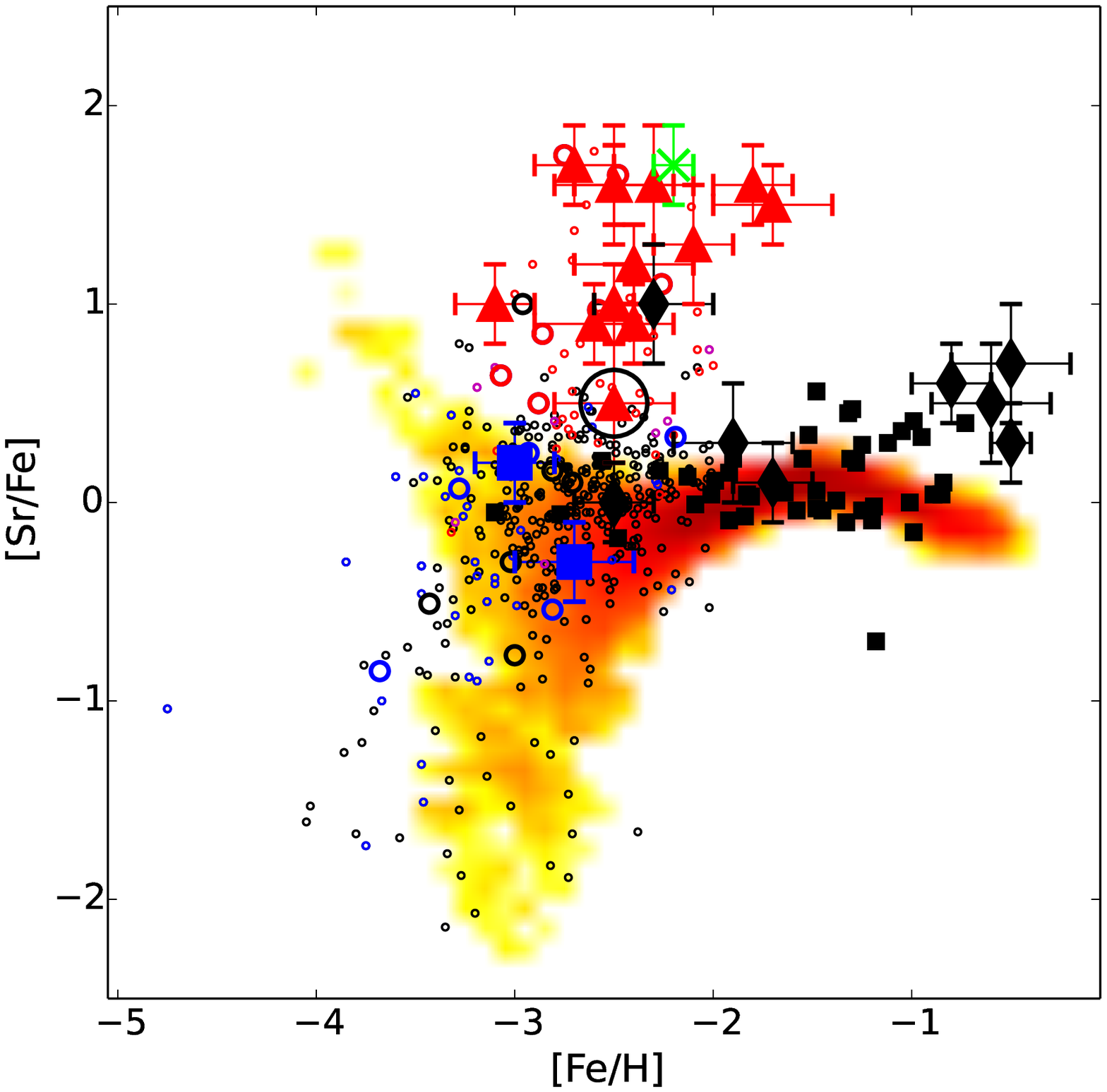}
\includegraphics[width=0.33\textwidth]{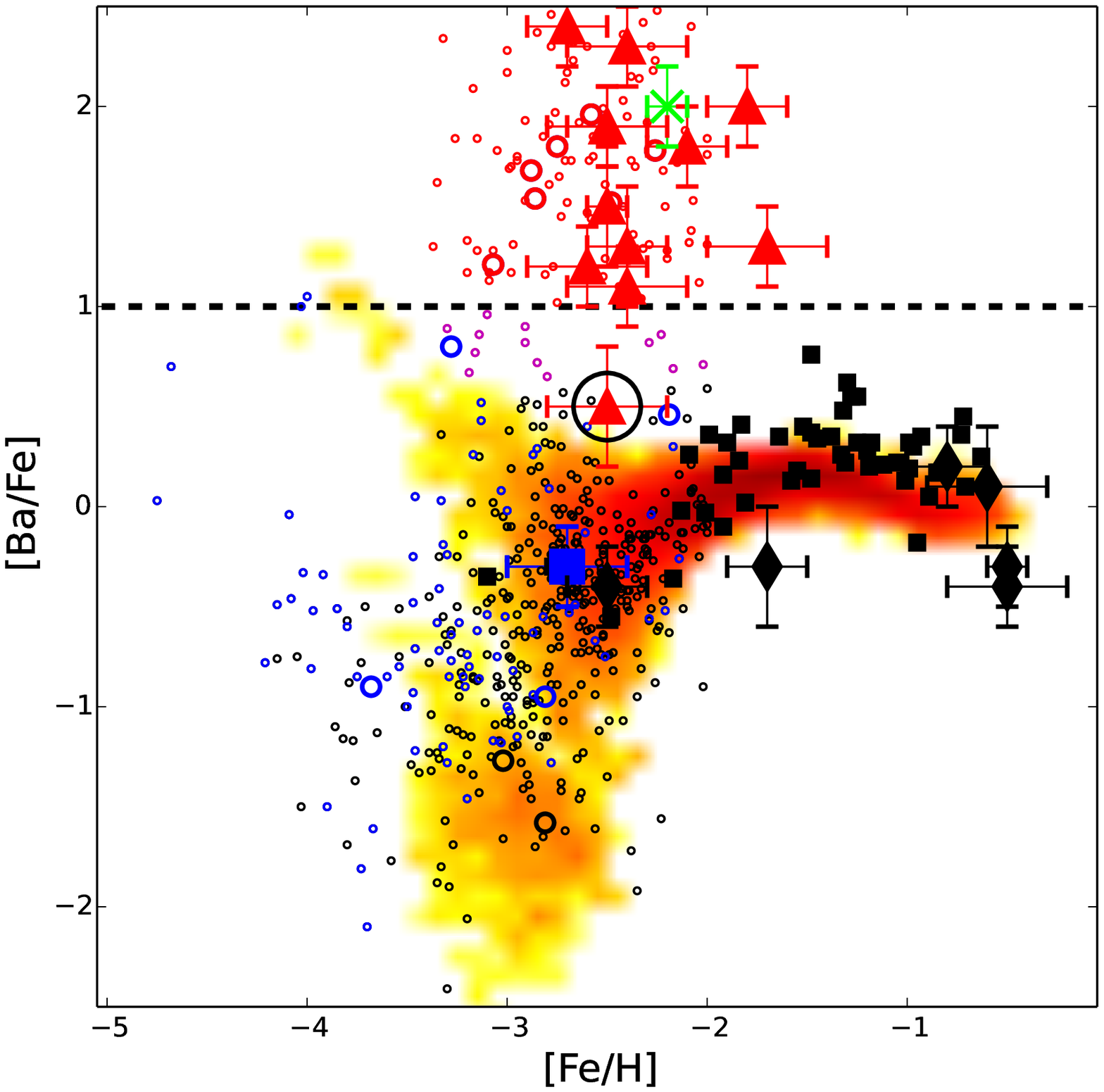}
\includegraphics[width=0.33\textwidth]{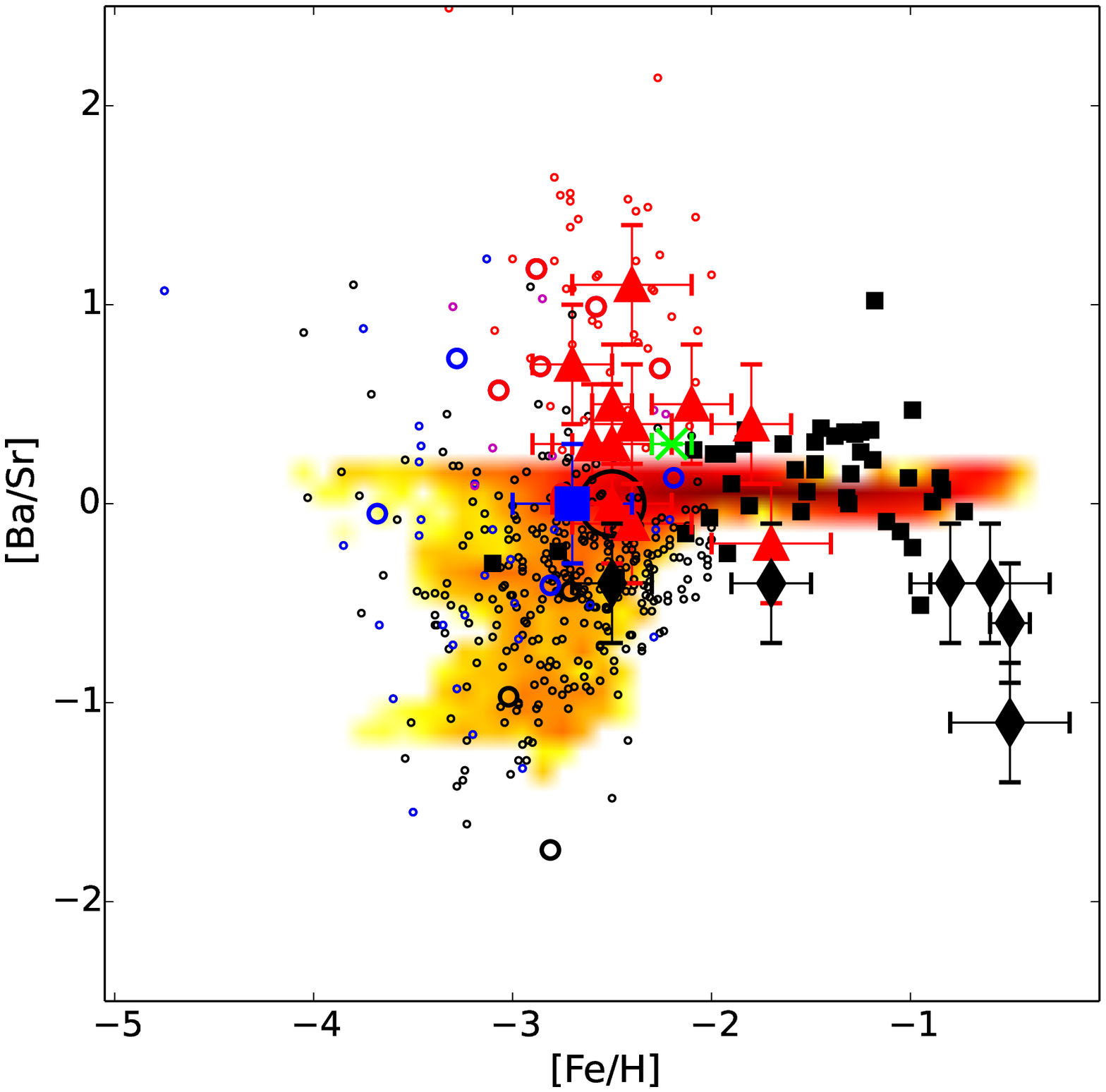}
\caption{Neutron-capture element abundances vs. [Fe/H] for our stars compared to 
metal-poor star samples from \citet[][black squares]{Hansen2012},
\citet[][large, open circles]{THansen2015}, and \citet{Placco2014b} and
references therein (small, open circles), respectively. Colours and symbols 
as in Fig.~\ref{comptolit}. The red-yellow cloud shows the GCE 
predictions, where the red colour indicates a larger number density of 
stars than the yellow. The dashed line at [Ba/Fe] = 1.0 in the lower panel 
separates carbon-normal from CEMP-$s$ stars. The bottom panel shows [Ba/Sr] vs. [Fe/H]. \label{GCE:SrBa}}
\end{center}
\end{figure}

\begin{figure}
\centering
\includegraphics[width=0.49\textwidth]{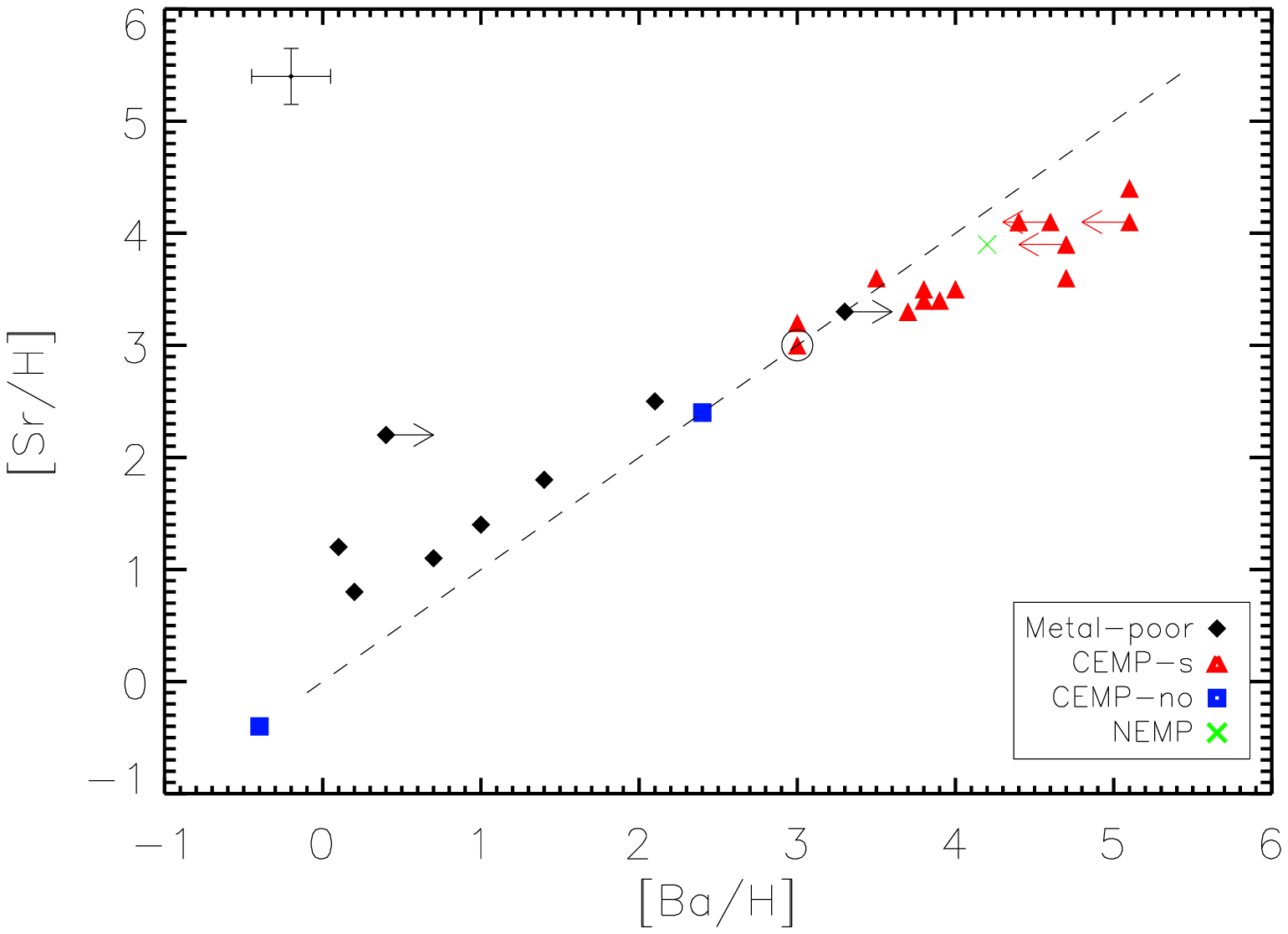}
\includegraphics[width=0.49\textwidth]{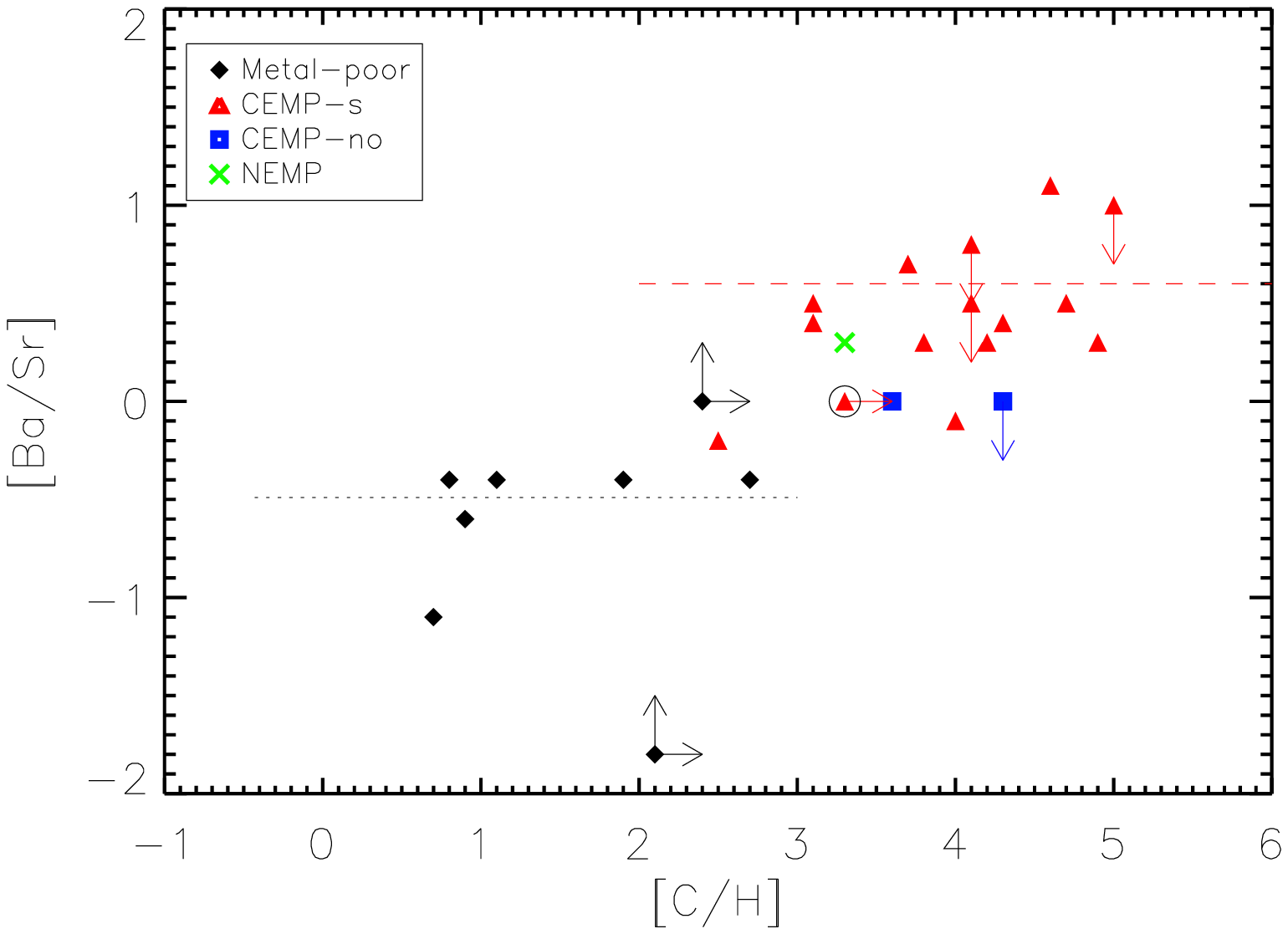}
\caption{Top: [Sr/H] as a function of [Ba/H], bottom: [Ba/Sr] as a
function of [C/H]. Note the offset in the trends between VMP and
the CEMP stars. Two stars in the bottom panel (one VMP and one CEMP-$s$, ) lie between the two levels 
at [C/H] $\sim 2.5$; these are the two variable AGB star candidates. 
The dashed and dotted lines indicate the average [Ba/Sr] ratios of the 
two groups of stars (see text). }
\label{Processes}
\end{figure}

Figure~\ref{GCE:SrBa} compares the X-shooter measurements of the
[Sr/Fe], [Ba/Fe], and [Ba/Sr] abundance ratios for our programme stars to larger
samples with high-resolution determinations from \citet{Hansen2012},
\citet{Placco2014b}, and \citet{THansen2015}. As seen from these
figures, the CEMP-no stars follow the general trend of stars without
C-enhancement predicted by standard Galactic chemical evolution (GCE) 
models, while the CEMP-$s$ stars
increase the star-to-star scatter, not just at extremely low
metallicities, but over the entire range of [Fe/H]. This is confirmed by
the Ba-Sr relation seen in Fig.~\ref{Processes}, a relation also found
by \citet{Roederer2013} for a much larger sample.

\begin{figure}
\centering
\includegraphics[width=0.49\textwidth]{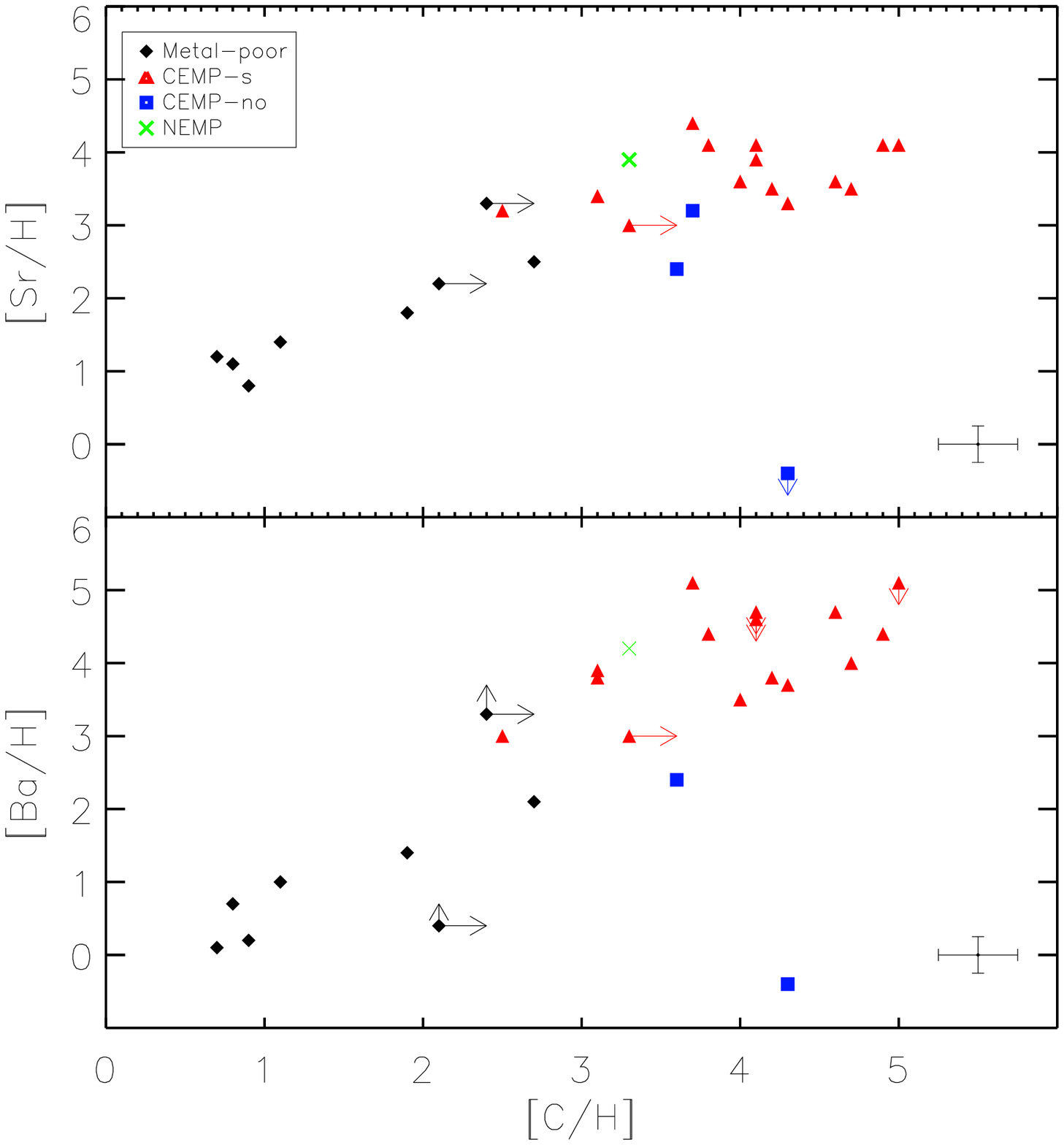}
\caption{[Sr/H] vs. [C/H] (top) and [Ba/H] vs. [C/H] (bottom). The outlier is \object{HE~2319$-$5228}.}
\label{SrHCH}
\end{figure}

Both VMP stars as well as CEMP stars exhibit a steady increase 
in their neutron-capture elements as a function of time (or [Fe/H]).
Figure~\ref{SrHCH} shows this trend very clearly, in terms of the absolute 
Sr, Ba, and C abundances. Iron has intentionally been excluded in these figures,
since Fe is formed in larger amounts by SNe of type Ia, which cannot explain 
the formation of CEMP-no (or CEMP-$s$) stars. A possible formation site of the early CEMP-no stars is faint core-collapse SN (type II), which might explode
with an O-Ne-Mg core only \citep[e.g.,][]{Wanajo2011}, indicating that they 
will not produce a significant amount of iron. Thus, iron (and its formation
processes) might confuse such trends, where we look for similarities in
abundance ratios to trace the underlying formation process.

Figure~\ref{Processes} shows a difference in the neutron-capture
element abundance trends between the C-normal VMP and CEMP-no stars
compared to the CEMP-$s$ stars. This would be expected if CEMP-$s$ stars
were enriched by a nucleosynthesis process that differs from that
reponsible for enriching metal-poor stars of similar [Fe/H], but lower
C-enhancement. The CEMP-$s$ stars are also shown to contain more Ba (a
main $s$-process elements) than Sr (a weak $s$-process element),
yielding a lower [Sr/Ba] ratio and pointing towards early AGB
enrichment in a binary system \citep{Lucatello2005,Masseron2010,
Cristallo2011, Bisterzo2012,Starkenburg2014}.

The CEMP-$s$ stars in the top panel of Fig.~\ref{Processes} fall
below the Ba-Sr 1:1 relation, while the VMP stars fall above
-- an indication of different formation scenarios that could be hidden
in the large star-to-star abundance scatter of Sr and Ba. The CEMP-no
stars fall in-between these two groups, right on the 1:1 relation. 

The ratio of Sr and Ba can be used to trace the formation
site -- AGB stars vs. spinstars or faint SN II -- and the [Ba/Sr] value
can even provide insight into the actual mass of the donor AGB
star. The predicted [hs/ls] ratio\footnote{The logarithmic abundance ratio 
of heavy-to-light [hs/ls] $s$-process elements, where we here use [Ba/Sr] as an
indicator.} from a metal-poor AGB star of 1.5-2$M_{\odot}$ is $\sim 0.5$,  according to the F.R.U.I.T.Y. database \citep{Cristallo2011}. Within the
observational uncertainty, this is in fair agreement with the observed
[Ba/Sr]$_{average} \sim 0.45$, based on all CEMP-$s$ stars in Fig.~\ref{Processes} (dashed red line). The C-normal VMP stars show the 
opposite trend ([Ba/Sr]$_{average} \sim -0.55$; dotted black line). This 
element ratio could be created by fast-rotating stars, which
might produce some Sr and little Ba, depending on conditions such as the
rotation velocity \citep{Frischknecht2012,Piersanti2013,Cristallo2015}. 

These relations are in good agreement with Galactic chemical evolution 
predictions (Fig.~\ref{GCE:SrBa}), where yields from spinstars may explain 
the chemically normal stars, but not stars with
extreme $s$-process enhancements \citep{Cescutti2013,Hansen2013}.

\subsection{Comparison to Galactic Chemical Evolution (GCE) models}

Figure~\ref{GCE:SrBa} shows the results from the GCE model presented in
\citet{Cescutti2013}. These results are based on the
stochastic enrichment of the ISM, assuming that the pollution by stars
are mixed only inside volumes with the radius of a typical SN II bubble
\citep[see][]{Cescutti2008}.  This model assumes $s$-process production
by spinstars for Sr and Ba, responsible mainly for the region of low
[Sr/Fe] and [Ba/Fe] at [Fe/H] $ < -2.0$, coupled with a production of 
neutron-capture elements by an $r$-process in electron-capture 
SNe. By combining
yields from these formation sites, the model can account for the
dispersion of the data in [Sr/Fe] and [Ba/Fe] for normal stars (and
CEMP-no stars)\footnote{We note that these results are independent of
other proposed $r$-process sites \citep[see, e.g.,][]{Cescutti2014}
combined with spinstars.  This was confirmed in \citet{Cescutti2015} using neutron star mergers as an $r-$process site. Similar results can be obtained considering
only two different primary ($r-$)process contributions not involving
spinstars \citep[see, e.g.,][]{Hansen2011,Hansen2014b}.}.

Fig.~\ref{GCE:SrBa} (bottom panel) highlights the importance of the production by spinstars of Sr and Ba
at low metallicity  coupled with an $r$-process production. 
In fact, not only rotating massive stars at low metallicity can produce 
a modest amount of Sr and Ba (compared to the production of an  $r$-process event),
but - even more interesting - this  $s$-process production  has an  [Ba/Sr] ratio varying between 0 and $-1.5$.
As the predictions of  the stochastic chemical evolution model show,
this represents also a  solution for the observed dispersion in [Ba/Sr] present 
in normal stars at extremely low metallicity.

As mentioned above, the chemical 
evolution model follows the Galactic halo ISM, and therefore cannot 
predict the enrichment observed in CEMP-$s$ stars. In any case, their 
atmospheres do not reflect the abundances of the
ISM where these stars are formed, since the original chemical
composition is altered by mass transfer from a binary
companion, the most likely formation scenario for the
majority of these objects. 

\subsection{Carbon isotopic ratios, $^{12}$C/$^{13}$C}

Convection, either in AGB stars or in spinstars, drives the CNO
cycle and transports $^{13}$C and N, created at the expense of $^{12}$C,
to a star's surface. This is detectable in the isotopic abundance ratios
of C and N, where low $^{12}$C/$^{13}$C and [C/N] ratios indicate strong
internal mixing with CN-cycled material \citep{Spite2005}. For 13
of our programme CEMP stars we find ratios that cover $3 < $
$^{12}$C/$^{13}$C $ < 50$, where the higher end of this range is in good
agreement with \citet{Bisterzo2011,Bisterzo2012}, who find that their
AGB models do not result in strongly mixed material 
($4 < ^{12}$C/$^{13}$C $ < 10$).

The equilibrium value of $^{12}$C/$^{13}$C for CNO-cycled material is
3-4, so stars with $^{12}$C/$^{13}$C = 5 indicate a high level of
processing. This either points towards more processing in the AGB stars
than current models predict, or it indicates processing by the star
itself.  Seven stars (\object{HE~0058$-$3449}, \object{HE~0241$-$3512}, \object{HE~0400$-$2030},
\object{HE~0414$-$0343}, \object{HE~0440$-$3426}, \object{HE~0448$-$4806}, \object{HE~2319$-$5228}) exhibit $^{12}$C/$^{13}$C ratios $<10$; two of these are subgiants and five are 
giants, in which some degree of internal processing is expected. We recall 
from Fig.~\ref{CNmixing} that \object{HE~0400$-$2030} and \object{HE~2319$-$5228} showed clear signs of mixing.

\subsubsection*{\object{HE~1238$-$0836} and \object{HE~2144$-$1832}}

Two stars stand out by having a  low log $g$ and a positive [C/N] 
ratio, namely \object{HE~1238$-$0836} and \object{HE~2144$-$1832}. Both 
of these stars are also photometric variables. 
According to \citet{Goswami2010}, \object{HE~1238$-$0836} is an RV Tau-type (R Sct) star; such stars can
exhibit extreme variations in their light curves.
As mentioned in Sect.~\ref{lit:RVsct}, we expect to have observed the
star in (or close to) an expansion phase, where the gravity and in turn
the pressure-sensitive abundances are too low. Multi-epoch
follow-up observations would be needed to verify the variable character.
We need to observe the star in a quiet phase, during which the
derived stellar parameters and abundances can be trusted. However, there
is also another possibility. Both stars could be pulsating, intrinsic
AGB stars such as \object{CS~30322-023} \citep{Masseron2006} and
\object{HD~112869} \citep{Zacs2015}. This would explain their low
gravities and the fact that both stars are bright (brighter than most
other stars in our programme). Either option could lead to [C/N]
$> 0$, and would most likely result in high $^{12}$C/$^{13}$C-ratios.

\subsection{CEMP-no and \emph{NEMP} stars \label{sec:CEMPno}}

To date we know $\sim 80$ CEMP-no stars \citep[most of which have
been included in][]{Placco2014b}\footnote{Many studies over the
past decades have steadily added to this sample, e.g.: 
\citet{Norris1997b, Christlieb2002,Frebel2005,Aoki2007,Norris2007,
Ito2013,Yong2013, Keller2014,Placco2014b,Bonifacio2015,THansen2015,Li2015}.}. Not all of these are confirmed CEMP-no stars,
but the majority are expected to belong to this sub-class. 
Our confirmation of three new CEMP-no stars (\object{HE~2155$-$2043},
\object{HE~2250$-$4229}, and \object{HE~2319$-$5228}) adds to this sample. 
The nature of the progenitors that enriched the ISM from which such stars 
formed (faint SNe with mixing and fallback, and/or spinstars) 
is still not fully understood. 

In the spinstars scenario, rotation triggers mixing
processes inside the star, and this leads to the production of important
quantities of primary $^{14}$N, $^{13}$C, and $^{22}$Ne compared to
stellar evolution models without rotation \citep{Hirschi2007}. However, 
this scenario does not fit the abundance pattern of one of the most-studied
CEMP-no stars, \object{BD+44$^{ \circ}$493} \citep{Ito2013,Placco2014a,
Maeder2015b}. \citet{Ito2013} found a nitrogen enhancement in this star
that is too low with respect to the carbon enhancement to match
these predictions. The abundance pattern of \object{HE~2319$-$5228}, however,
appears to be consistent with a spinstar progenitor that may have been
present in the earliest generations of stars \citep[e.g.,
][]{Maeder2015}.

\subsubsection*{The NEMP star: \object{HE~0400$-$2030}}

\citet{Johnson2007} defined the nitrogen-enhanced metal-poor (NEMP)
stars as objects with [C/N] $\la -0.5$ and [N/Fe] $ > 0.5$, and found
five examples. This number was lower than expected, and seemed to point
towards an observational bias against discovering NEMP stars relative to
CEMP stars (which are more readily identifiable from objective-prism
surveys). However, the number of recognized NEMP stars has increased to
$\sim 45$ at present \citep{Placco2014b}, which is a substantial
number, although still less than the number of known CEMP-no and CEMP-$s$ stars. 
We find one new NEMP star: \object{HE~0400$-$2030}, which has [C/N]$ = -1.4$ and 
is also $s-$process-element enhanced.

\subsection{Comparison to AGB yields}

Due to hot-bottom-burning (HBB), the higher-mass stars (of 4 $-$ 6\,
$M_{\odot}$) produce prodigious amounts of nitrogen at the expense of
carbon. A few CEMP stars studied to date exhibit abundance patterns that
could be consistent with these higher-mass AGB stars, but the stars in
our X-shooter sample appear to be largely consistent with the 1.5--3
$M_{\odot}$ cases, where nitrogen is not enhanced relative to carbon, as
would be expected from a HBB scenario. 

Among the seven stars that show indications of mixing through their
[C/N] ratios or isotopic carbon-abundance ratios, none exhibit
[C/N] $< -0.5$ and [hs/ls] $ > 0.5$ simultaneously. This indicates that
HBB in more massive stars ($\ga4\,M_{\odot}$) is a less plausible
explanation for the chemical composition of these stars. Most of our
programme stars exhibit signs of weak internal mixing, which, based on 
their [Ba/Sr] ([hs/ls]) ratios, can be explained by either AGB stars, 
massive spinstars, or faint SNe with mixing and fallback.

\section {Conclusions}

Twenty-seven stars from our X-shooter programme were analysed through
spectral synthesis of molecular C and N bands, as well as atomic Ba
and Sr lines. With only these four abundances, it is possible to
classify each star according to its abundance pattern (i.e., CEMP-$s$, 
CEMP-no etc.). The majority of the known CEMP stars are enriched in 
$s$-process elements such as Ba and Sr. These CEMP-$s$ stars appear to
belong to the relatively metal-rich inner-halo population, while the
CEMP-no stars may belong primarily to the relatively more metal-poor
outer-halo population \citep[e.g.,][]{Carollo2012,Carollo2014}. 

Despite intense efforts to date, we are still trying to understand the
exact formation sites of the CEMP-no and NEMP stars. Two of our newly confirmed
CEMP-no stars appear to fall on (or slightly above) the low-C band
suggested by \citet{Spite2013}. We also confirm the differences 
found between the strongly C-enhanced CEMP-$s$ stars and the
relatively less C-enhanced CEMP-no stars discussed by the same authors. 
 However, several stars in this and other larger studies appear to indicate a continuum of absolute C abundances, rather than
discrete bands. Here we show that differences in the \emph{heavy} element abundances as a function of
the absolute carbon abundances ([C/H]) show similar plateau trends for the two sub-classes
of CEMP stars. We note that larger samples may erase these plateaus and exhibit more continouos distributions around similar average values.

Comparison of the CEMP-$s$ stars to AGB model yields 
\citep[e.g.,][]{Cristallo2011} indicates that the
progenitor AGB stars were primarily of the lower-mass variety (in
agreement with \citealt{Kennedy2011} and \citealt{Bisterzo2012}), while the NEMP
star in our programme could be associated with a more massive AGB progenitor
 capable of producing large abundances of nitrogen relative to carbon. However, the [hs/ls]
ratio for this star agrees with values predicted for
lower mass AGB stars ($\sim 2\,M_{\odot}$).  The two stars with low
gravities (\object{HE~1238$-$0836} and \object{HE~2144$-$1832})
appear to be pulsating variables, which
could yield more trustworthy abundances if observed in a quiet
phase. They could also be (intrinsic) AGB stars.

This study has shown that moderate-resolution, low-S/N X-shooter
spectra are of sufficient quality to classify CEMP stars into
sub-catogories and extract information on their plausible astrophysical
formation sites. This highlights an aspect of the X-shooter instrument 
that we consider very promising, 
the ability to obtain simultaneous measurements of C, N, Ba, and Sr very efficiently for a relatively large sample of faint
stars. This is important, in particular for intermediate to low-metallicity
$-2.5 < $ [Fe/H] $ < -1.0$ stars, a range we cover with our
sample. This metallicity interval has been largely ignored in recent
observational campaigns that concentrate on the most extreme metal-poor
stars. Substantial and crucial information can be extracted at intermediate
metallicity, where numerous stars of the halo system are found. For example,
different $r$-process sites predict dispersions in stellar abundances of
[Sr/H] or [Ba/H], as discussed by \citet{Cescutti2014} and
\citet{Hansen2012,Hansen2014b}.


\begin{acknowledgements}

C.J.H. acknowledges support from research grant VKR023371 from the
Villum Foundation, and both she and T.T.H. acknowledge support from
Sonderforschungsbereich SFB 881. "The Milky Way System" (subproject A5)
of the German Research Foundation (DFG). B.N. and C.J.H. thank Drs. J.
Fynbo and D. Malesani for help with the X-shooter observations. 
B.N. acknowledges partial support by the National Science
Foundation under Grant No. NSF PHY11-25915. T.T.H. thanks T. Masseron for line list information. 
T.C.B., C.R.K., and V.M.P.
acknowledge partial funding of this work from grants PHY 08-22648;
Physics Frontier Center/Joint Institute or Nuclear Astrophysics (JINA),
and PHY 14-30152; Physics Frontier Center/JINA Center for the Evolution
of the Elements (JINA-CEE), awarded by the US National Science
Foundation. J.A. and B.N. acknowledge support from the Danish Council
for Independent Research | Natural Sciences and the Carlsberg Foundation.

This publication has made use of the SIMBAD database, operated at CDS,
Strasbourg, France, and of data products from the Two Micron All Sky
Survey, which is a joint project of the University of Massachusetts and
the Infrared Processing and Analysis Center/California Institute of
Technology, funded by the National Aeronautics and Space Administration
and the National Science Foundation.

\end{acknowledgements}
\bibliographystyle{aa}

\end{document}